\documentclass[aps,amsmath,twocolumn,amssymb,floatfixng,showpacs,
superscriptaddress,footinbib]{revtex4-1}

\usepackage{bm}
\usepackage{graphicx}
\usepackage{graphics}
\usepackage{mathtools}
\usepackage{amsmath}
\usepackage{amsfonts}
\usepackage{amssymb}
\usepackage{epstopdf}
\usepackage{hyperref}
\usepackage{hyperref}
\hypersetup{
    colorlinks,%
    citecolor=blue,%
    linkcolor=blue,%
    urlcolor=blue
}
\usepackage[normalem]{ulem}
\usepackage{tikz}\usetikzlibrary{petri}

\newcommand{\be}   {\begin{equation}}
\newcommand{\ee}   {\end{equation}}
\newcommand{\ba}   {\begin{eqnarray}}
\newcommand{\ea}   {\end{eqnarray}}

\newcommand{\bs}{\boldsymbol}

\newcommand{\customSection}[1]{{{\it{#1.}}---}}

\begin{document}

\title{Competing magnetic states on the surface of multilayer ABC-stacked graphene}
\author{Lauro B. Braz}
\affiliation{
Instituto de F\'{\i}sica, Universidade de S\~ao Paulo, Rua do Mat\~ao 1371, S\~ao Paulo, S\~ao Paulo 05508-090, Brazil
}
\affiliation{
 Department of Physics and Astronomy, Uppsala University, Box 516, S-751 20 Uppsala, Sweden
}
\author{Tanay Nag}
\affiliation{
 Department of Physics and Astronomy, Uppsala University, Box 516, S-751 20 Uppsala, Sweden
}
\affiliation{Department of Physics, BITS Pilani-Hyderabad Campus, Telangana 500078, India}

\author{Annica M. Black-Schaffer}
\affiliation{
 Department of Physics and Astronomy, Uppsala University, Box 516, S-751 20 Uppsala, Sweden
}

\date{ \today }

\begin{abstract}
We study interaction-mediated magnetism on the surface of ABC-multilayer graphene driven by its zero-energy topological flat bands. Using the random-phase approximation we treat onsite Hubbard repulsion and find multiple competing magnetic states, due to both intra- and inter-valley scattering, with the latter causing an enlarged magnetic unit cell.
At half-filling and when the Hubbard repulsion is weak, we observe two different ferromagnetic orders.
Once the Hubbard repulsion becomes more realistic, new ferrimagnetic orders arise with distinct incommensurate intra- or inter-valley scattering vectors depending on interaction strength and doping, leading to a multitude of competing magnetic states.

\end{abstract} 

\maketitle


Graphene has extraordinary electronic and structural properties \cite{geimRiseGraphene2007,Neto09,novoselovElectricFieldEffect2004}. 
While ideal graphene does not exhibit magnetic properties, 
 various derivatives of graphene do \cite{Palacios07,Young-Woo06,yazyevEmergenceMagnetismGraphene2010,sharpeEmergentFerromagnetismThreequarters2019b},
e.g.~in the presence of a sublattice imbalance \cite{Esquinazi03,Palacios08,magdaRoomtemperatureMagneticOrder2014,Lemonik12} or due to Landau levels \cite{McCann06,Nomura06,Goerbig11,youngTunableSymmetryBreaking2014} or interlayer twists  \cite{bistritzerMoireBandsTwisted2011a,sharpeEmergentFerromagnetismThreequarters2019b,chenTunableCorrelatedChern2020,Klebl21,Gonzalez-Arraga17,Xiao18,Jianpeng19}. In particular, modifications generating flat bands with a high density of states (DOS) close to the Fermi energy are prone to not only magnetism, but generally correlated insulators and even superconductivity \cite{caoUnconventionalSuperconductivityMagicangle2018a,luSuperconductorsOrbitalMagnets2019,xieSpectroscopicSignaturesManybody2019,poOriginMottInsulating2018a,liObservationVanHove2010a,sharpeEmergentFerromagnetismThreequarters2019a,choiElectronicCorrelationsTwisted2019b,caoCorrelatedInsulatorBehaviour2018,jiangChargeOrderBroken2019a,andrei2020graphene,Tomas17}. The flat band dispersion quells kinetic energy and thus even weak electron-electron interaction generates electronic ordering.


In monolayer graphene, the large Fermi velocity generated by the linear $k$-dispersion prevents electronic ordering. 
Bilayer AB-stacked (Bernal) and trilayer  ABC-stacked (rhombohedral) graphene host quadratic $k^2$ and cubic $k^3$ band dispersions, respectively, resulting in electronic instabilities, with bilayer \cite{velascoTransportSpectroscopySymmetrybroken2012,geisenhofQuantumAnomalousHall2021,Gonzalez-Arraga_17,zhouIsospinMagnetismSpinpolarized2022a,pangburnSuperconductivityMonolayerFewlayer2023a,crepieuxSuperconductivityMonolayerFewlayer2023a,pantaleonSuperconductivityCorrelatedPhases2023b,marchenkoExtremelyFlatBand2018}, trilayer \cite{leeCompetitionSpontaneousSymmetry2014,yankowitzElectricFieldControl2014,pangburnSuperconductivityMonolayerFewlayer2023a,crepieuxSuperconductivityMonolayerFewlayer2023a,zhouSuperconductivityRhombohedralTrilayer2021b}, and also tetralayer graphene \cite{myhroLargeTunableIntrinsic2018,kerelskyMoirelessCorrelationsABCA2021} having been found to exhibit magnetic, as well as superconducting ordering, but so far only with application of electric and/or magnetic fields \cite{Avetisyan09,Min07,Gava09,Koshino10,luiObservationElectricallyTunable2011,MacDonald10}.
Further stacking of $N$ layers in an ABC-sequence produces a locally flat $k^N$ band dispersion on the surfaces of the stack, protected by topology \cite{heikkilaFlatBandsTopological2011a,heikkilaDimensionalCrossoverTopological2011a,mcclureElectronEnergyBand1969b,Henck18}. 
These flat surface states have been shown to host versatile Fermi surface properties \cite{Hongki08,MacDonald10,Koshino10,Yuan11,baoStackingdependentBandGap2011,Jeil13,luiObservationElectricallyTunable2011,gaoLargeareaEpitaxialGrowth2020,kaladzhyanSurfaceStatesQuasiparticle2021} leading to magnetism \cite{shiElectronicPhaseSeparation2020b,hagymasiObservationCompetingCorrelated2022c,pamukMagneticGapOpening2017f,henniRhombohedralMultilayerGraphene2016a,leeGateTunableMagnetismGiant2022a,otaniIntrinsicMagneticMoment2010a,xiaoDensityFunctionalInvestigation2011a,cuongMagneticstateTuningRhombohedral2012a,pamukMagneticGapOpening2017f,hanCorrelatedInsulatorChern2024}
and theory proposals also exist for superconductivity, without any need for external fields \cite{lothmanUniversalPhaseDiagrams2017d,awogaSuperconductivityMagnetismSurface2023b,heikkilaFlatBandsTopological2011a,kopninHightemperatureSurfaceSuperconductivity2011b}.

Theoretical understanding of ABC-stacked multilayer graphene (ABC-MLG) \cite{MacDonald10,Macdonald2011} {even predating  experimental findings,} is primarily based on work, including first-principles calculations, studying only the primitive (in-plane) unit cell, then finding surface ferrimagnetism; 
opposite magnetic moments on the two sublattices with one moment substantially suppressed \cite{Otani10,Dong-Hui12,Olsen13,Pamuk17,Ruijuan11,cuongMagneticstateTuningRhombohedral2012a, awogaSuperconductivityMagnetismSurface2023b}. {Recent experiments have further reported domain formation \cite{shiElectronicPhaseSeparation2020b}, interpreted as a competition between a ferrimagnetic state and a suggested correlated paramagnetic state, while longer-range magnetic ordering has only been discussed for finite doping \cite{hagymasiObservationCompetingCorrelated2022c}}. 
In this work we study the formation of magnetic ordering on the surface of ABC-MLG, in particular, taking into account all possible magnetic ordering patterns.

We use the $T$-matrix formalism to isolate the ABC-MLG surface Green's function for $N\gg1$ layers and then incorporate electronic interactions in the form of Hubbard on-site repulsion $U$ within the matrix random-phase approximation (RPA). 
We find strong ordering tendencies for both intra- and inter-valley scattering, generating single and extended unit cell magnetic patterns, respectively.
At half-filling, we find putative ferromagnetic (FM) ordering centered on one sublattice only and mediated by the non-interacting flat bands, but  is quickly suppressed for small interactions $U$.
At more realistic $U$-values we find opposite but unequal moments on the two sublattices, rendering ferrimagnetic (FiM) ordering at incommensurate scattering vectors. The inter-valley ordering requires lower $U$ for ordering, while the intra-valley ordering is close to the commensurate ferrimagnetic order reported in earlier work \cite{Otani10,Dong-Hui12,Olsen13,Pamuk17,Ruijuan11,cuongMagneticstateTuningRhombohedral2012a, awogaSuperconductivityMagnetismSurface2023b,hagymasiObservationCompetingCorrelated2022c}. Adding finite doping, intra-valley ordering instead requires the lowest $U$, with a substantially shorter spin-spin relaxation time.
Our work establishes a fierce competition between different magnetic orders, as well as the importance of incommensurability.  



\customSection{Model and method}
{To model ABC-MLG we consider a tight-binding model of the bulk unit cell in the basis $\left\{ c_{\tilde{\bs{k}}}^{A_1},\, c_{\tilde{\bs{k}}}^{B_1},\, c_{\tilde{\bs{k}}}^{A_2},\, c_{\tilde{\bs{k}}}^{B_2},\, c_{\tilde{\bs{k}}}^{A_3},\, c_{\tilde{\bs{k}}}^{B_3} \right\}$, where $A_n$($B_n$) denotes sublattice A(B) in layer $n$ \cite{kaladzhyanSurfaceStatesQuasiparticle2021,Jung13}.
We use intra(inter)-layer hopping $\gamma_{1}=3.3$~eV($\gamma_{2}=0.42$~eV) between $A_n \to B_n$ ($B_n \to A_{n+1}$), while finite inter-layer hopping $\gamma_{3}$ between $A_n \to B_{n+1}$ is responsible for trigonal warping, splitting the graphene Dirac cone into three satellite Dirac cones causing a triangular Fermi surface \cite{Jung13}, using both $\gamma_3/\gamma_1 =0,0.1$ for unwarped and warped ABC-MLG. \textcolor{black}{We note that intralayer next nearest neighbor hopping is redundant as satellite Dirac cones already exist.}
We further use intralayer nearest neighbor distance $a_{0}=1.42$~Å, and interlayer nearest neighbor distance $d_{0}=2.36a_{0}$ and vary the chemical potential through $\mu$.}

{As we are interested in the ABC-MLG surface in the limit of $N\gg1$ layers, we introduce a virtual wall of impurities separating the bulk system into two semi-infinite pieces along the $z$-direction. 
Following the $T$-matrix formalism, we construct the {\it surface} Green's function $\hat{G}(\tilde{\boldsymbol{k}}_{1},\tilde{\boldsymbol{k}}_{2},i\omega_{n})$ from the bulk Green's function $\hat{G}_0(\tilde{\boldsymbol{k}},i\omega_{n})$ \cite{Pinon20,kaladzhyanSurfaceStatesQuasiparticle2021}, 
with $\omega_{n}$ the Matsubara frequency.
We extract the surface Green's function of one of the surfaces adjacent to the impurity plane, $\hat{\mathcal{G}}(\boldsymbol{k},i\omega_{n})$, by partial Fourier transform of $\hat{G}(\tilde{\boldsymbol{k}}_{1},\tilde{\boldsymbol{k}}_{2},i\omega_{n})$ in the $z$-direction, using henceforth the in-plane notation $\boldsymbol{k}=(k_x,k_y)$. 
This results in an effective surface Green's function for a three layer deep, six carbon atom, single surface unit cell, see schematic in Fig.~\ref{fig1}(d). \textcolor{black}{This exact numerical method allows us to encode the effects of a large number of internal layers into $\hat{G}(\tilde{\boldsymbol{k}}_{1},\tilde{\boldsymbol{k}}_{2},i\omega_{n})$}
\footnote{We verify that the surface spectral function $A(\bs{k},\omega)=-{\rm Im}[ {\rm Tr}[\sum_{i\omega_n}\hat{\mathcal{G}}(\boldsymbol{k},z_{1},z_{2},i\omega_{n}+\omega)]/\pi]$, reproduces the surface flat bands \cite{kaladzhyanSurfaceStatesQuasiparticle2021}}. For more details, see further Supplementary Material (SM) \cite{supp,ozakiContinuedFractionRepresentation2007b}.

{As a realistic, yet minimal, model of} electron-electron interactions, we consider on-site intra-orbital Hubbard repulsion $U\sum_{ia}n^a_{i\uparrow}n^a_{i\downarrow}$ where $n^a_{i\sigma} = c^{a\dagger}_{i\sigma}c^a_{i\sigma}$ is the occupation number for site $i$, orbital $a$, and spin $\sigma$. {This interaction is not only the largest term in graphene \cite{wehlingStrengthEffectiveCoulomb2011c}, but nearest-neighbor repulsion is also irrelevant in ABC-MLG since each flat band only occupies one sublattice \cite{otaniIntrinsicMagneticMoment2010a,cuongMagneticstateTuningRhombohedral2012a,hagymasiObservationCompetingCorrelated2022c}. It also reproduces first-principles calculations (of single unit cells) that include the full range of Coulomb interactions \cite{lothmanUniversalPhaseDiagrams2017d,awogaSuperconductivityMagnetismSurface2023b}.}

We track the influence of interactions on both charge and magnetic fluctuations by employing the RPA to extract the relevant susceptibilities. Magnetic phenomena are governed by the spin susceptibility matrix given by \cite{scalapinoWavePairingSpindensitywave1986c,graserNeardegeneracySeveralPairing2009b,kemperSensitivitySuperconductingState2010a}
\begin{align}
\hat{\chi}_{\text{s}}(\boldsymbol{q},\omega) & =\hat{\chi}_{0}(\boldsymbol{q},i\omega)\left[\hat{1}-\hat{U}_{s}\hat{\chi}_{0}(\boldsymbol{q},i\omega)\right]^{-1}, \label{eq:spin_RPA_matrix} \\
\chi_{td}^{sp}(\boldsymbol{q},\omega)&=-\frac{1}{\beta N}\sum_{\boldsymbol{k},i\omega_{n}}\mathcal{G}_{st}(\boldsymbol{k},i\omega_{n})\mathcal{G}_{pd}(\boldsymbol{k}+\boldsymbol{q},i\omega_{n}+i\omega).\label{eq:non-interacting}
\end{align}
where the non-interacting, or bare, susceptibility matrix $\hat{\chi}_{0}(\boldsymbol{q},\omega)$ has size $N_{b}^{2}\times N_{b}^{2}$ and is constructed from the surface non-interacting Green's function. 
$N_{b}=6$ denotes the number of carbon sites {in the surface unit cell,} indexed as $s,p,t,d=A_1,~\cdots, B_3$, while ${\bs q}=(q_x,q_y)$ accounts for the scattering momentum,  and $\beta=\left(k_{B}T\right)^{-1}$ is the inverse temperature. The spin interaction matrix $\hat{U}_{s}$ takes the simple form $U\delta_{sp} \delta_{pt} \delta_{td}$ \cite{graserNeardegeneracySeveralPairing2009b}. 
To extract magnetically ordered states we use the density-density correlation functions,  namely the homogeneous spin susceptibility \cite{mahanManyParticlePhysics2000,bruusManyBodyQuantumTheory2004}, also experimentally tractable  \cite{kreiselTheorySpinExcitationAnisotropy2022},
\begin{equation}
\chi(\boldsymbol{q},\omega)=\frac{1}{2}\sum_{s,p}\chi_{pp}^{ss}(\boldsymbol{q},\omega).
\label{eq:homogeneous}
\end{equation}
In particular, we compute the real [imaginary] parts of the homogeneous contributions for both the bare susceptibility, $\chi^{\prime}_0(\boldsymbol{q},\omega)$ [$\chi^{\prime\prime}_0(\boldsymbol{q},\omega)$], and the RPA susceptibility $\chi^{\prime}_{\text{s}}(\boldsymbol{q},\omega)$ [$\chi^{\prime\prime}_{\text{s}}(\boldsymbol{q},\omega)$]. In a similar treatment for the charge susceptibility, we find that it always remains small compared to the spin susceptibility. We thus conclude that charge fluctuations are not important in ABC-MLG.

A divergent RPA susceptibility signals the formation of an ordered state, for magnetism expressed by the (generalized) Stoner criterion $\text{det}[\widetilde{U}_{s}\hat{\chi}_{0}(\boldsymbol{q},\omega)- \lambda \hat{1}]  =0$ with $\widetilde{U}_{s}=\hat{U}_{s}/U$ \cite{sakakibaraOriginMaterialDependence2012a}. 
We thus obtain the critical Hubbard interaction strength for magnetic ordering from the maximum positive eigenvalue $\lambda_{s}^{{\bs q}_m}={\rm max}\{\lambda^{\bs q}_s \}=1/U^{{\bs q}_m}_{\rm c}$. We perform this analysis over the full first Brillouin zone (BZ) to capture longer-range magnetic ordering patterns, beyond single-unit cell ordering, to include all incommensurate ordering. 
The eigenvector $\bs{\psi}^{\omega=0}_{\bs {q}_m}$, associated with the eigenvalue $\lambda_{s}^{{\bs q}_m}$ once the Stoner criterion is satisfied, encodes the spatial structure of the magnetic order \cite{Fong2000,Gao2010,boehnkeSusceptibilitiesMaterialsMultiple2015,Christensen16,fischerUnconventionalSuperconductivityMagicangle2022}.
Furthermore, a resonance structure in the dynamic profile of $\chi^{\prime\prime}_s(\boldsymbol{q}_m,\omega)$ at specific $\omega=\tilde{\omega}$ in the ordered regime, i.e.~for $U>U^{{\bs q}_m}_{\rm c}$, can be assigned as the spin gap of the underlying order \cite{Fong2000,Gao2010,Christensen16}.
{We further remark that all magnetic states  in this work, i.e.~within one surface, fall within the layer antiferromagnet (LAF) phase \cite{Macdonald2011} when considering the two opposite surfaces of any slab,  see SM \cite{supp}.}

\customSection{Flat bands and nesting}
We start by analyzing the non-interacting surface states of ABC-MLG. Fig.~\ref{fig1}(a) shows how the surface spectral function $A(\bs{k},\omega=0)$ acquires substantial weight, signaling a flat band area, around the $\bs{k}=\bs{K},\tilde{\bs{K}}$ points
in a tripartite and triangular shape due to the three satellite Dirac points \cite{Jung13}, due to the trigonal warping $\gamma_3$. If instead $\gamma_3=0$, a circular shape is instead achieved, see Fig.~\ref{fig1}(b). Moving away from half-filling, the surface spectral function takes the shape of an annular ring, see Fig.~\ref{fig1}(c), now instead capturing the bulk (dispersive) Dirac cones. The surface unit cell with impurity wall is represented in Fig. ~\ref{fig1}(d).

\begin{figure}[ht]
\begin{center}
\includegraphics[width=1\linewidth]{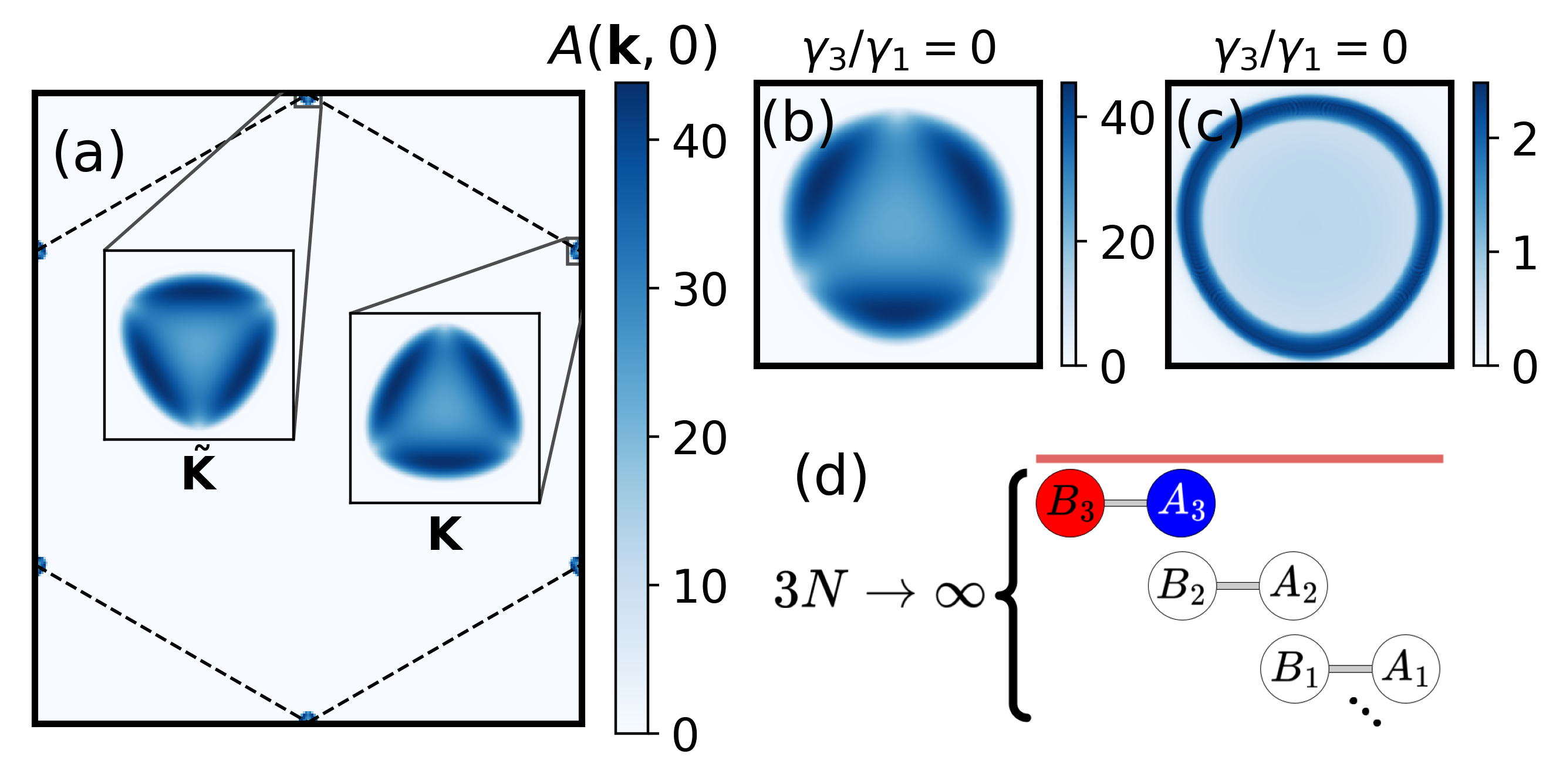}
\end{center}
\vspace{-17pt}
\caption{Surface spectral function $A(\bs{k},\omega=0)$ in the first BZ displaying triangular Fermi surfaces around $\bs{k}\approx \bs{K},\tilde{\bs{K}}$ for $\gamma_3/\gamma_1=0.1$ (a). Structure of  $A(\bs{k}\approx \bs{K},\omega=0)$ for $\mu=0$~eV (b) and $\mu=0.025$~eV at $\gamma_3/\gamma_1=0$ (c).  {
Surface unit cell with sites $A_1 \cdots B_3$ in three layers with impurity wall (red line) (d)}.}
\label{fig1}
\end{figure}

The existence of two Fermi surfaces, around $\bs{K}$ and $\tilde{\bs{K}}$, leads to both intra- and inter-valley scattering with scattering, or nesting, vectors $\bs{q} \approx \bs{\Gamma},\bs{K}$, respectively. This suggests that more than one type of magnetic order, with different $U_{\rm c}$'s, may be present.
Before investigating magnetic ordering driven by finite interactions, we analyze the effect of nesting on the bare spin susceptibility. 
In Fig.~\ref{fig2}(a) we plot the real static non-interacting susceptibility  $\chi^{\prime}_0(\boldsymbol{q},\omega=0)$ at $\mu=0$ and note large values in small regions ${\bs \Gamma}_{-} <\bs{q}< {\bs \Gamma}_{+} $ and  ${\bs K}_{-} <\bs{q}<{\bs K}_{+}$, with divergences at the commensurate nesting vectors $\bs{q} ={\bs \Gamma},{\bs K}$. The finite region are due to the finite extent of the surface flat bands.
Away from half-filling, $\chi^{\prime}_0(\boldsymbol{q},\omega=0.025)$ instead shows only finite peaks in the same regions, see Fig.~\ref{fig2}(b). Thus divergences in the bare susceptibility are only present due to the zero-energy flat bands and limited to half-filling. This follows directly from the standard representation of the bare homogeneous 
susceptibility $\chi_0=\sum_{l,p}\chi^{ll}_{pp,0} (\bs {q}) \propto \sum_{\bs k}(f(E_{l}^{\boldsymbol{k}+\boldsymbol{q}})-f(E_{p}^{\boldsymbol{k}}))/(\omega+E_{l}^{\boldsymbol{k}+\boldsymbol{q}}-E_{p}^{\boldsymbol{k}}+i0^{+})$, with $f(E)$ the Fermi-Dirac distribution, where the energy denominator vanishes for the flat bands. We further find no notable dependence on the trigonal warping $\gamma_3$.

With a divergent bare spin susceptibility, even an infinitesimal small interaction $U$ triggers magnetic ordering at half-filling according to the generalized Stoner criteria. Analyzing the resulting eigenvectors at ${\bs \Gamma}$ and ${\bs K}$, we find magnetic moments only on the top surface's $B_3$ orbital. 
For the ${\bs K}$-ordering the magnetic moment is also modulated in real space, acquiring zero net magnetization. 
Due to this magnetic structure, we refer to both of these states as sublattice ferromagnetic (sFM) order and note that they originate from commensurate nesting vectors.
To gain more understanding of the sFM orders, we examine the imaginary part of the dynamic bare spin susceptibility $\chi^{\prime \prime}_{s}(\bs{q},\omega)$ at the divergent scattering vectors ${\bs q}= {\bs \Gamma},{\bs K}$ in Figs.~\ref{fig2}(b,d) as a function of small $U$. 
We find a large (negative) peak starting from $(\omega > 0,U=0)$ and ending at $(\omega= 0,U=U^{\pm}_{\rm c})$ with $U^{-}_{\rm c} \approx 0.5$~eV for ${\bs q}= {\bs \Gamma}$ and $U^{+}_{\rm c} \approx 1.0$~eV for ${\bs q}= {\bs K}$.  
This indicates the existence of a finite spin gap protecting the sFM orders, but only for $U<U^{\pm}_{\rm c}$. 
Although the $\bs K$-sFM state has zero magnetization and absence of time-reversal symmetry, the characteristic two-resonance structure of chiral magnons in altermagnets \cite{maierWeakcouplingTheoryNeutron2023,smejkalChiralMagnonsAltermagnetic2023} is not seen, so this order is unlikely to emerge.
For a more in-depth discussion on the sFM orders and their susceptibilities, see SM \cite{supp}.

\begin{figure}[t]
\begin{center}
\includegraphics[width=0.98\linewidth]{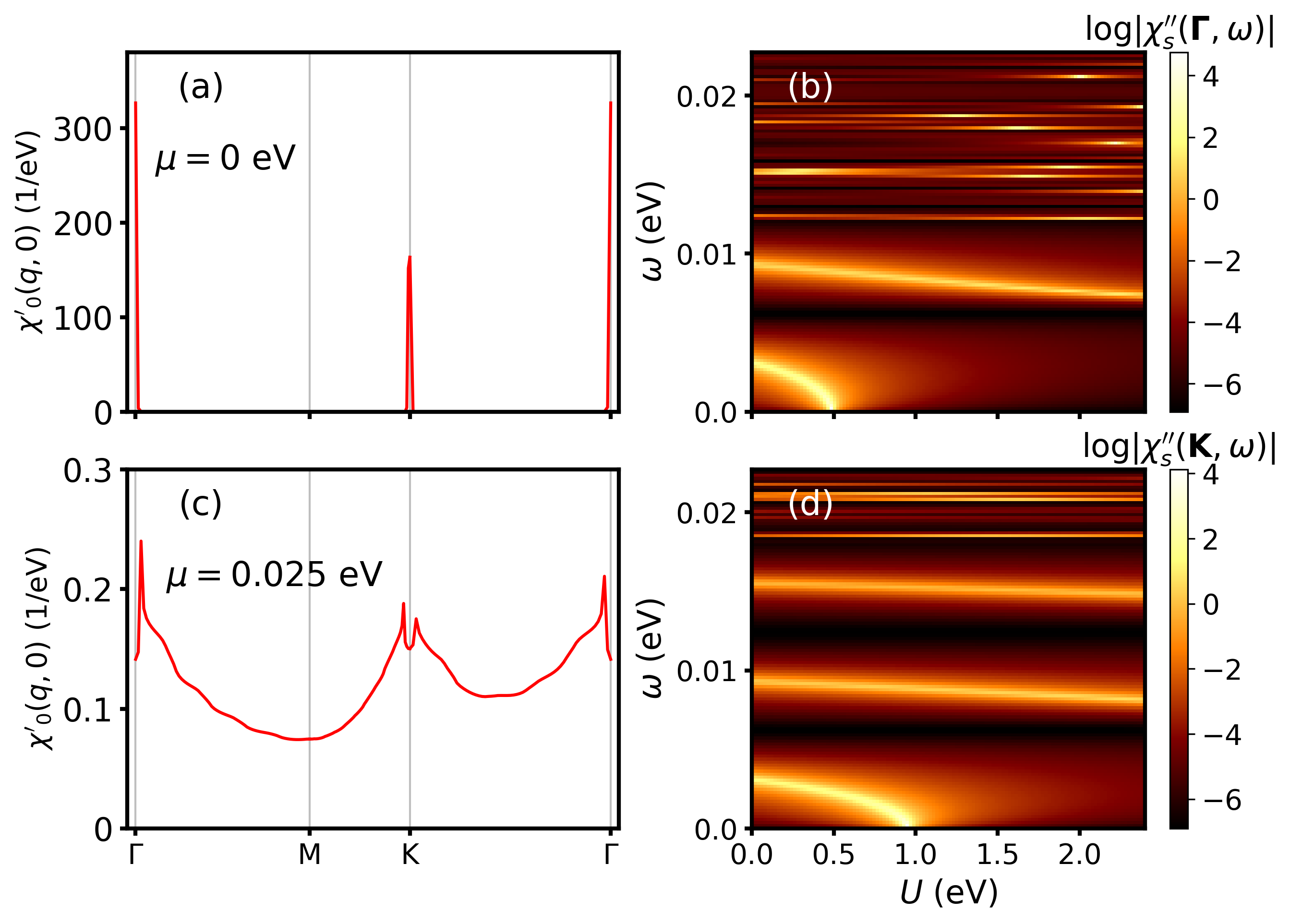}
\end{center}
\vspace{-17pt}
\caption{Real part of the bare spin susceptibility $\chi^{\prime}_{0}(\bs{q},0)$ at $\mu=0$~eV (a) and $\mu=\omega=0.025$eV (c) along high-symmetry BZ directions. 
Imaginary part of the dynamic homogeneous susceptibility $\chi^{\prime\prime}_{0}(\bs{q},\omega)$ in logarithmic scale
in the $\omega$-$U$ plane for ${\bs q}={\bs \Gamma}$ (b) and ${\bs K}$ (d) for $\mu=0$~eV. Here $\gamma_3/\gamma_1=0.1$.
}
\label{fig2}
\end{figure}

\customSection{FiM order with interactions}
With $U^{\pm}_{\rm c} \ll \gamma_1$ and estimations of the on-site repulsion in graphene and graphite instead giving $U\gtrsim \gamma_1$ \cite{wehlingStrengthEffectiveCoulomb2011c,tancogne-dejeanParameterfreeHybridlikeFunctional2020a,bursillOptimalParametrisationPariser1998a,vergesFitPariserParrPopleHubbard2010a}, we do not expect the sFM solutions to likely exist in ABC-MLG. 
We thus continue analyzing the spin susceptibility $\chi_{s}(\bs{q},\omega)$ for more realistic $U$, here first at half-filling. We find that as $U$ increases beyond $U^{+}_{\rm c}$, the static spin susceptibility $\chi^{\prime}_{s}(\bs{q},\omega=0)$ becomes substantially suppressed at both $\bs{q} = \bs{\Gamma},{\bs{K}}$.
We instead observe new divergences appearing for inter-valley scattering at $\bs{q}=\bs{K}'\approx\bs{K}_{-}$ with $U^{K'}_{\rm c} \approx 2.30$~eV and for intra-valley scattering at $\bs{q}=\bs{\Gamma}' \approx \bs{\Gamma}_{+}$ for $U^{\Gamma'}_{\rm c} \approx 4.40$~eV, see Figs.~\ref{fig3}(a,c). 
Both divergences appear at incommensurate scattering vectors, linked to the finite extent of the surface flat band and thus different from the non-interacting commensurate sFM states. 
We can qualitatively understand this shift in scattering vectors by noting that a finite interaction shifts the quasiparticle energies $E_{l}^{\boldsymbol{k}} \to E_{l}^{\boldsymbol{k}} + \Sigma_l^{\boldsymbol{k}}$ with a self-energy $\Sigma_l^{\boldsymbol{k}}$ \cite{Berk66}, such that the momentum dependence of $\chi_{s}$ may be different from that of $\chi_{\rm 0}$.
We further note that the peak in $\chi^{\prime}_{s}(\bs{q},\omega=0)$ changes from positive to negative values as soon as the Stoner criterion is satisfied at $U^{\Gamma',K'}_{\rm c}$ \cite{Christensen16,Fong2000,Gao2010}. 
The negatively valued divergency is necessary due to the previous magnetic transitions at $U < U^{\Gamma',K'}_{\rm c}$ found in the previous section, see SM \cite{supp}.

In terms of the resulting magnetic moments for the ${\bs \Gamma'}$ state, we find that the dominant contribution to $\chi^{\prime}_{s}(\bs{q},0)$ comes from the top surface $B_3$ orbital and now also with the $A_3$ orbital carrying an unequal finite magnetic moment but with opposite sign, as illustrated in the Fig.~\ref{fig3}(d) inset (incommensurability not visible).
A similar analysis of the ${\bs K'}$ state results in the pattern in the Fig.~\ref{fig3}(b) inset, again with dominant contribution from a $B_3$ orbital, but now with a $\sqrt{3} \times \sqrt{3}$ extended spatial repetition, due to the inter-valley scattering \cite{boehnkeSusceptibilitiesMaterialsMultiple2015,Fong2000,Gao2010}, see SM \cite{supp}. 
{Due to these patterns, we refer to the $\bs{\Gamma}'$ and $\bs{K}'$ states as incommensurate ferrimagnetic (FiM) orders, but note that the latter can also be called an incommensurate spin density wave.}

Moreover, we consider the dynamic profile of $\chi^{\prime \prime}_{s}(\bs{q},\omega)$ in Fig.~\ref{fig3}(b,d), now at the incommensurate wave vectors $\bs{q}=\bs{\Gamma}',\bs{K}'$ hosting the divergent (real) spin susceptibility \footnote{The change to incommensurate wave vectors means the peaks in Fig.~\ref{fig2}(b,c) at small $U$ for commensurate vectors are not visible in Fig.~\ref{fig3}(b,d).}. 
We find a large positive peak originating at $U = U^{\Gamma',K'}_{\rm c}$ and $\omega =0$  and rising to larger frequencies $\omega$ with increasing $U$.
This signals the existence of finite spin gaps \cite{bruusManyBodyQuantumTheory2004,kreiselTheorySpinExcitationAnisotropy2022}, confirming the formation of first a $\bs{K}'$-FiM state at $U^{K'}_{\rm c}$ from inter-valley scattering and then a $\bs{\Gamma}'$-FiM ordered state at $U^{\Gamma'}_{\rm c}$ from intra-valley scattering.
With the spin susceptibility at $\bs{K}'$ becoming substantially suppressed at $U^{\Gamma'}_{\rm c}$, see Fig.~\ref{fig3}(c), we infer that the $\bs{\Gamma}'$-FiM order likely directly set in at $U>U^{\Gamma'}_{\rm c}$ with little competition from the $\bs{K}'$-FiM order, see SM \cite{supp}.
We further find that $\bs {\Gamma'},\bs{K}'$ varies slightly with trigonal warping, producing slightly different $U_c$'s, but not changing the overall behavior.
Taken together, these results point to a close competition between the ${\bs \Gamma'}$-FiM and ${\bs K'}$-FiM states at half-filling, such that any spatial dependence of $U$ will likely create spontaneous domains of different FiM states. 
%
 
\begin{figure}[t]
\begin{center}
\includegraphics[width=0.98\linewidth]{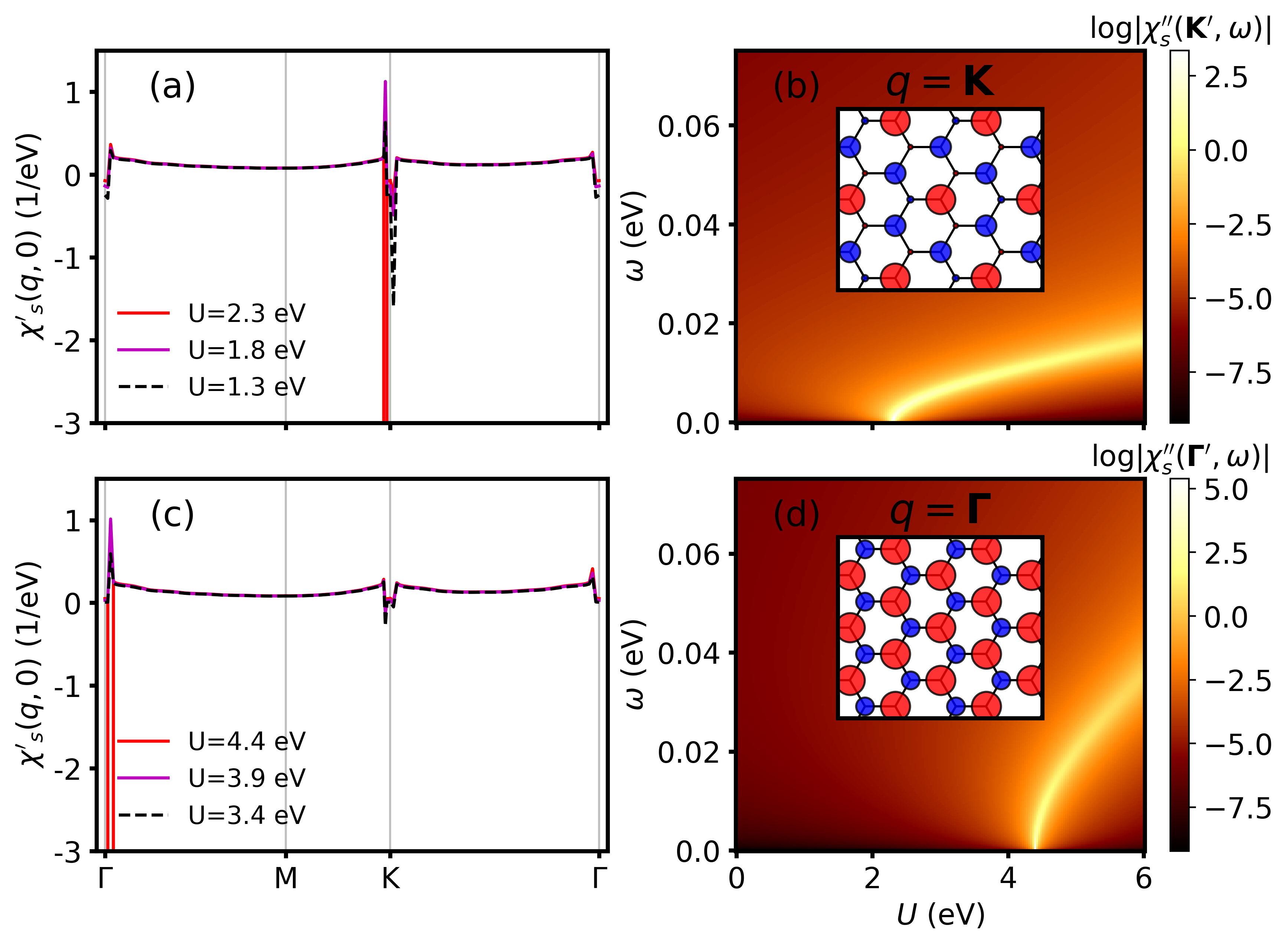}
\end{center}
\vspace{-17pt}
\caption{ 
Real part of the spin susceptibility $\chi^{\prime}_{s}(\bs{q},0)$  for $U^{+}_{\rm c}<U \lesssim U^{K'}_{\rm c}$ (a) and $U^{K'}_{\rm c}<U \lesssim U^{\Gamma'}_{\rm c} $ (c) along high-symmetry BZ directions.  
Imaginary part of the dynamic homogeneous susceptibility $\chi^{\prime\prime}_{0}(\bs{q},\omega)$ in logarithmic scale
in the $\omega$-$U$ plane for $\bs{q}=\bs{\Gamma}'$ (b) and $\bs{q}=\bs{K}'$ (d).
Insets: Real space magnetic structure associated with $\bs{q}= \bs{K}'$-, and $\bs{\Gamma}'$-FiM states on top surface layer (incommensurability ignored). Circle radii are determined by overall magnet moment magnitudes, colors differentiate signs.
Here $\mu=0$ and $\gamma_3/\gamma_1=0.1$. 
}
\label{fig3}
\end{figure}

\customSection{Finite doping} 
Finally, we vary the doping away from half-filling, modeling spontaneous charge inhomogeneity \cite{hagymasiObservationCompetingCorrelated2022c} or an applied gate voltage \cite{shiElectronicPhaseSeparation2020b,hagymasiObservationCompetingCorrelated2022c}.
To probe the doping dependence, we extract the minimum critical Hubbard parameter $U^*_c$ at zero temperature satisfying the Stoner criterion for any nesting vector $\bs{q}$ in discs centered at $\bs{\Gamma},\bs{K}$ to capture all previously explored divergences and beyond. 
In Fig.~\ref{fig6} we plot the result as a function of doping $\mu$ {(lower-horizontal axis) with  associated number density $n_e$ (upper-horizontal axis)}. We find increasing $U^*_c$ for both intra-valley (red) and inter-valley scattering (blue), with higher $U^*_c$ for the latter. 
At $\mu=0$~eV the results coincide with Fig.~\ref{fig2} with its commensurate orders already at  $U^*_c = 0$. 
With increasing doping, we find a roughly linear increase in $U^*_c$, while at the same time, the ordering vectors move away from $\bs{\Gamma},\bs{K}$. 
The resulting scattering vectors are labeled $\bs{\Gamma}''$ and $\bs{K}''$, and the order as $\bs{\Gamma}''$- and $\bs{K}''$-FiM orders as they show large similarities with the $\mu=0$ FiM orders, see SM \cite{supp}. 

We plot the deviation $\Delta q_{\Gamma}=|\bs{\Gamma}''-\bs{\Gamma}|$ and similar for $\bs{K}$ in the inset of Fig.~\ref{fig6}. There is a sharp jump in $\Delta q$ directly when $\mu$ acquires finite value, followed by a slow increase for increasing $\mu$. 
We find that $\Delta q_{\Gamma,K}$ fully tracks the position of the finite amplitude peaks in the bare susceptibility, see Fig.~\ref{fig2}(c) for a fixed $\mu$. Thus magnetic ordering at finite doping is fully determined by the bare susceptibility with the interaction just enhancing its peaks into divergences at $U_c^*$. 
This is notably different from the half-filled case where the interaction also shifts the ordering vectors to incommensurate vectors, compare Figs.~\ref{fig2}(a) and \ref{fig3}(a,c).

\begin{figure}[t]
\begin{center}
\includegraphics[width=0.98\linewidth]{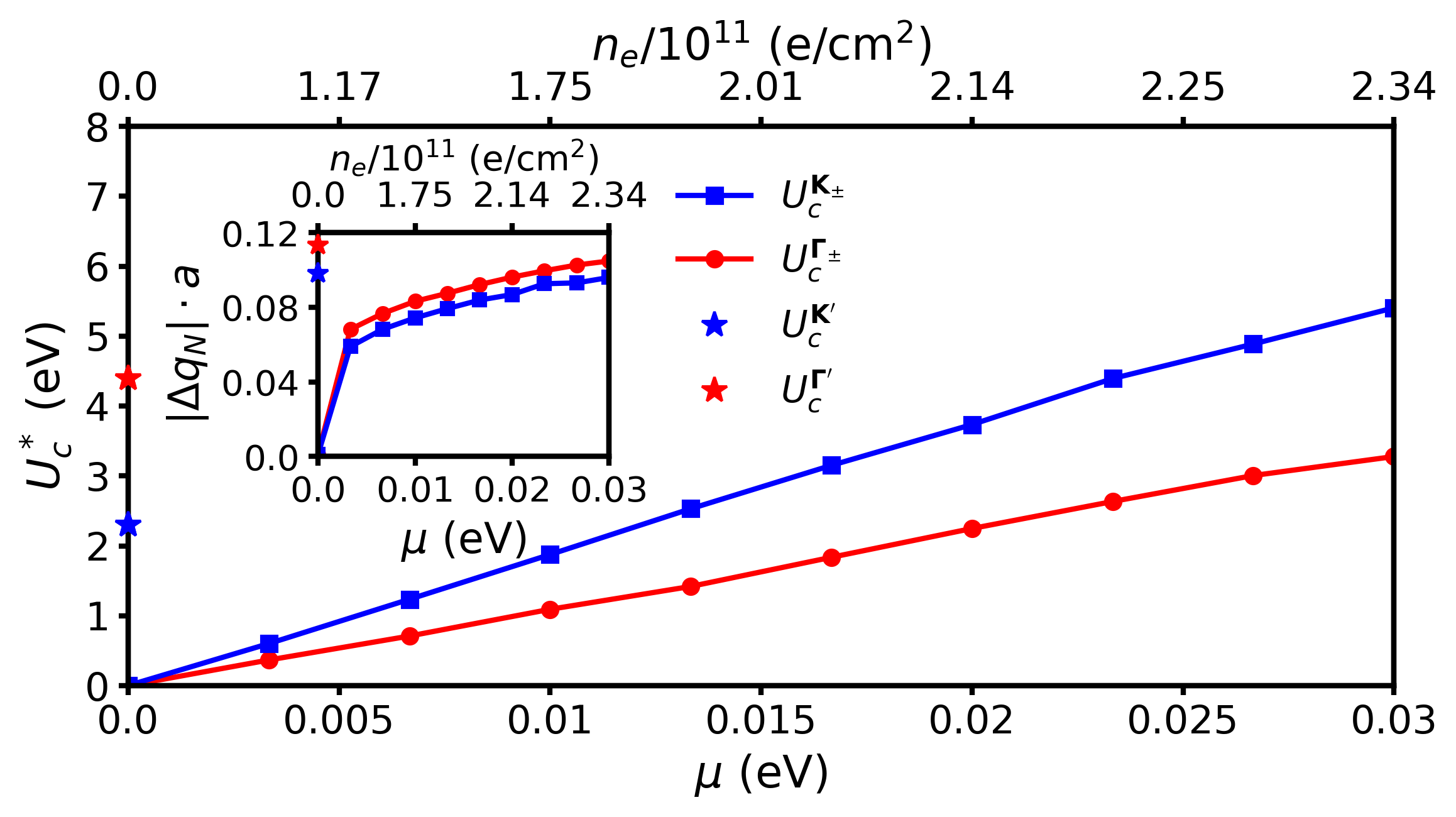}
\end{center}
\vspace{-17pt}
\caption{Minimum critical Hubbard interaction $U^*_c$ extracted for all scattering vectors in a circular region around $\bs{\Gamma}$ (red) and $\bs{K}$ (blue) as a function of doping $\mu$ and carrier density $n_e$. (Inset) Deviation in the ordering vectors with respect to the high-symmetry points $\Delta\bs{q}_N={\bs \Gamma}''({\bs K}'')-{\bs \Gamma}({\bs K})$ normalized by the real-space graphene lattice constant $a$ as a function of $\mu$ and carrier density $n_e$.
Stars indicate $U_c^{\Gamma'},U_c^{K'}$ and deviation for ${\bs \Gamma}'$, ${\bs K}'$ in main plot and inset, respectively. {Corresponding number density (units of $e/{\rm cm}^{2}$) on the upper horizontal axes (not linear scale).}
Here $\gamma_3/\gamma_1 = 0.1$.
}
\label{fig6}
\end{figure}

To highlight further differences to the half-filling results in Fig.~\ref{fig3}, we mark $U_c^{\Gamma',K'}$ with stars in Fig.~\ref{fig6}. 
As seen, at half-filling inter-valley FiM order commands the lowest $U_c$, while at any finite doping where intra-valley scattering has the lowest $U_c$. 
Also, the incommensurability is always largest at half-filling, which seems to provide an upper limit for $\Delta {\bs q}$ at finite doping.
We also find a notably different overall behavior of the spin susceptibility. 
At half-filling a clear spin gap is present at $U=0$, protecting the commensurate sFM orders until its closure at $U_c^{\pm}$, while at $U_c^{\Gamma',K'}$, a new spin gap develops protecting the FiM orders as shown in Figs.~\ref{fig2}(b,d) and \ref{fig3}(b,d).
Instead, at finite doping no spin gap, or order, exists for any $U<U_c^*$. 
Moreover, when ordering is established at $U_c^*$, we find no sharp peaks at finite frequencies in $\chi_s''$. 
The dynamic susceptibility displays signatures of a markedly short spin-spin relaxation time \cite{barraElectronSpinResonance2005,yalcinFerromagneticResonance2013,maierWeakcouplingTheoryNeutron2023},
flattening the peaks in $\chi''$ and also making $\chi'$ continuous at non-zero frequencies. 
We attribute the shortened relaxation time to more substantial overlap with the dispersive surface states at finite doping \cite{maierWeakcouplingTheoryNeutron2023}, generating damping.
Interestingly, the bulk metallic states do not generate any strong damping, as obvious from orders at half-filling.
Still tracing a spin gap from the (flattened) peaks in $\chi_s''$, we find that they never disappear with increasing $U$, once developed at $U_c^*$. For more information, see SM \cite{supp}. 
Based on the different $U_c$s and their trends, scattering vectors, and different relaxation times, we conclude that the states at finite doping are notably different from those at half-filling.

\customSection{Concluding remarks}
Our work reveals how the surface flat band in ABC-multilayer graphene leads to a plethora of magnetic states: at half-filling, commensurate sFM orders with either intra- and inter-valley nesting vectors in the non-interacting limit, but at any realistic interaction strength, incommensurate FiM orders instead develop for both intra- and inter-valley scattering. 
Away from half-filling, interactions result in another set of incommensurate FiM orders with different nesting vectors and a shorter spin-spin relaxation times. 
This demonstrates remarkably rich possibilities for different magnetic ordering on the surface ABC-MLG: both competing intra- and inter-valley scattering and with ordering varying intricately as a function of interaction and doping \cite{maierWeakcouplingTheoryNeutron2023,smejkalChiralMagnonsAltermagnetic2023}. 

{We note that there exist qualitative similarities between our results on ABC-MLG with $N\gg1$ layers and recent experiments on few-layer graphene. For example, all our magnetic states belong to the layer antiferromagnet (LAF) phase recently found in pentalayer ABC-stacked graphene \cite{Macdonald2011,hanCorrelatedInsulatorChern2024}, but we additionally uncover a rich in-surface structure for these LAF states.
Also, in contrast to bilayer \cite{velascoTransportSpectroscopySymmetrybroken2012,geisenhofQuantumAnomalousHall2021,zhouIsospinMagnetismSpinpolarized2022a,marchenkoExtremelyFlatBand2018}, trilayer \cite{leeCompetitionSpontaneousSymmetry2014,yankowitzElectricFieldControl2014,zhouSuperconductivityRhombohedralTrilayer2021b}, tetralayer \cite{myhroLargeTunableIntrinsic2018,kerelskyMoirelessCorrelationsABCA2021}, and pentalayer \cite{hanCorrelatedInsulatorChern2024} graphene, external displacement field is not required to stabilize magnetic states in ABC-MLG.
The reduced spin-spin relaxation time with finite doping also agrees with the loss of magnetic ordering in fewer-layer systems.}
{Further, recent experimental work on ABC-MLG has reported a gapped magnetic order at half-filling and an enlarged magnetic unit cell at finite doping, similar to our $\bs{\Gamma}'$- and $\bs{K}'$-FiM states, respectively, but without incommensurability, together with an ungapped phase modelled as a correlated paramagnetic phase \cite{hagymasiObservationCompetingCorrelated2022c}.
Importantly, we find that both incommensurability and magnetic pattern vary intricately with interaction strength and doping, resulting in a multitude of energetically close but different domains in real samples. This likely results in increased quantum fluctuations and reduced energy gaps, consistent with current experiments \cite{shiElectronicPhaseSeparation2020b,hagymasiObservationCompetingCorrelated2022c}}.

We thank G.~B.~Martins for insights on the RPA method,  P.~Holmvall for providing his code implementation of the Matsubara summations, and X.~Feng, D.~Chakraborty, R.~Arouca, P.~M.~Oppeneer, and P.~Thundstr\"om for fruitful discussions. We acknowledge financial support from the Knut and Alice Wallenberg Foundation through the Wallenberg Academy Fellows program and project grant KAW 2019.0068, Coordenação de Aperfeiçoamento de Pessoal de Nível Superior - Brasil (CAPES) - Finance Code 001, and grant 2023/14902-8, São Paulo Research Foundation (FAPESP).

\bibliography{references,ABC-MLG_v5,ABC-MLG_v6}


\normalsize\clearpage
\begin{onecolumngrid}
	\begin{center}
		{\fontsize{12}{12}\selectfont
			\textbf{Supplementary Material for ``Competing magnetic states on the surface of multilayer ABC-stacked graphene''\\[5mm]}}
		{\normalsize Lauro B. Braz,$^{1,2}$  Tanay Nag,$^{2,3}$ and Annica Black-Schaffer,$^{2}$\\[1mm]}
		{\small $^1$\textit{Instituto de F\'{\i}sica, Universidade de S\~ao Paulo, Rua do Mat\~ao 1371, S\~ao Paulo, S\~ao Paulo 05508-090, Brazil}\\[0.5mm]}
		{\small $^2$\textit{Department of Physics and Astronomy, Uppsala University, Box 516, 75120 Uppsala, Sweden}\\[0.5mm]}
		{\small $^3$\textit{Department of Physics, BITS Pilani-Hyderabad Campus, Telangana 500078, India}\\[0.5mm]}
		{}
	\end{center}
	
	\newcounter{defcounter}
	\setcounter{defcounter}{0}
	\setcounter{equation}{0}
	\renewcommand{\theequation}{S\arabic{equation}}
	\setcounter{figure}{0}
	\renewcommand{\thefigure}{S\arabic{figure}}
	\setcounter{page}{1}
	\pagenumbering{roman}
	
	\renewcommand{\thesection}{S\arabic{section}}



This Supplementary Material (SM) provides additional information supporting the results in the main text. 
In Section~\ref{Sec:SI-1} we demonstrate the details of our model discussed in the section ``\textit{Model and method}" in the main text. In particular, we elaborate on the methodology used to exactly access the surface states in the $N\rightarrow\infty$ layer limit, and also how this implies that all our states fall within the so-called layer-antiferromagnetic (LAF) phase.
In Section~\ref{Sec:SI0} we establish where in the Brillouin zone divergences in the spin-susceptibility appear. This warrants the choices of reciprocal coordinates in the figures in the main text.
In Section~\ref{Sec:SI1b} we provide additional data on the dynamic spin susceptibility in the non-interacting case and at low interactions at half-filling, supplementing the results in the section ``\textit{Flat band and nesting effects}" in the main text.
In Section~\ref{Sec:SI1} we provide additional data on the dynamic spin susceptibility for realistic $U$ interaction strengths at half-filling, supplementing the results in the section ``\textit{FiM order with interaction}" in the main text.
In Section~\ref{Sec:SI2} we provide additional data on the dynamic spin susceptibility at finite doping, supplementing the results in the section ``\textit{Finite doping}" in the main text.
In Section~\ref{Sec:SI4} we demonstrate additional analysis on the magnetic spin texture, supplementing the magnetic structures provided in section ``\textit{FiM order with interactions}" in the main text.

\section{Surface Green's function formalism in the $N\rightarrow\infty$ layer limit and LAF state}
\label{Sec:SI-1}

{In Sec.~\textit{Model and method} we briefly describe the procedure used to access the surface Green's function in the $N\rightarrow\infty$ layer limit of ABC-MLG that we study in this work.
Here, we give more details on this method. 
To model ABC-MLG we start by considering a tight-binding model of bulk ABC-graphite in the basis $\left\{ c_{\tilde{\bs{k}}}^{A_1},\, c_{\tilde{\bs{k}}}^{B_1},\, c_{\tilde{\bs{k}}}^{A_2},\, c_{\tilde{\bs{k}}}^{B_2},\, c_{\tilde{\bs{k}}}^{A_3},\, c_{\tilde{\bs{k}}}^{B_3} \right\}$, where $A_n$($B_n$) denotes sublattice A(B) in layer $n=1,2,3$ in a three-layer unit cell: \cite{kaladzhyanSurfaceStatesQuasiparticle2021}:}
{\begin{equation}
\label{eq:Hbulk}
\begin{split}
&\mathcal{H}(\tilde{\bs{k}})=  \gamma_1h_1(\tilde{\bs{k}})[c_{\tilde{\boldsymbol{k}}}^{A_1 \dagger}c^{B_1}_{\tilde{\boldsymbol{k}}} + 
c_{\tilde{\boldsymbol{k}}}^{A_2 \dagger}c^{B_2}_{\tilde{\boldsymbol{k}}}+ 
c_{\tilde{\boldsymbol{k}}}^{A_3 \dagger}c^{B_3}_{\tilde{\boldsymbol{k}}}] +\gamma_3h_{3}(\tilde{\boldsymbol{k}})[c_{\tilde{\boldsymbol{k}}}^{A_1 \dagger}c^{B_2}_{\tilde{\boldsymbol{k}}}+c_{\tilde{\boldsymbol{k}}}^{A_2 \dagger}c^{B_3}_{\tilde{\boldsymbol{k}}}]+ \alpha f(\tilde{\boldsymbol{k}}) [\gamma_{2} c_{\tilde{\boldsymbol{k}}}^{B_3\dagger}c^{A_1}_{\tilde{\boldsymbol{k}}}+  \gamma_3h_{3}(\tilde{\boldsymbol{k}})   c_{\tilde{\boldsymbol{k}}}^{A_3 \dagger}c^{B_1}_{\tilde{\boldsymbol{k}}}] \\
&+\gamma_{2}c_{\tilde{\boldsymbol{k}}}^{B_1\dagger}c^{A_2}_{\tilde{\boldsymbol{k}}}+\gamma_{2}c_{\tilde{\boldsymbol{k}}}^{B_2\dagger}c^{A_3}_{\tilde{\boldsymbol{k}}}-
\mu\sum_{a}c_{\tilde{\boldsymbol{k}}}^{a \dagger}c^{a}_{\tilde{\boldsymbol{k}}}+\text{H.c.},
\end{split}
\end{equation}}
{where 
$h_{1}(\tilde{\boldsymbol{k}}) =1+2\exp(-3ia_{0}\tilde{k}_{x}/2)\cos(\sqrt{3} a_{0}\tilde{k}_{y}/2)$, 
$h_{3}(\tilde{\boldsymbol{k}}) =2\exp(-3ia_{0}\tilde{k}_{x}/2)\cos(\sqrt{3} a_{0}\tilde{k}_{y}/2)+\exp(-3ia_{0}\tilde{k}_{x})$, 
$f(\tilde{\boldsymbol{k}})=\exp(3ia_{0}\tilde{k}_{x})\exp(3id_{0}\tilde{k}_{z})$, with $\tilde{\boldsymbol{k}}=(\tilde{k}_{x},\tilde{k}_{y},\tilde{k}_{z})$, and $a=\{A_1,B_1,...,B_3,C_3\}$ is enumerating the orbitals (sites/carbon atoms). Here
$\gamma_{1}$($\gamma_{2}$) denotes the intra(inter)-layer hopping between $A_n \to B_n$ ($B_n \to A_{n+1}$), while a finite inter-layer hopping $\gamma_{3}$ between $A_n \to B_{n+1}$ is responsible for trigonal warping, splitting the graphene Dirac cone into three satellite Dirac cones causing a triangular Fermi surface \cite{Jung13}. By using $\alpha=1$  $(0)$ we capture bulk ABC-stacked graphite (quasi-2D ABC-trilayer graphene).  
We set $\gamma_{1}=3.3$~eV, $\gamma_{2}=0.42$~eV, intralayer nearest neighbor distance $a_{0}=1.42$~Å, and interlayer nearest neighbor distance $d_{0}=2.36a_{0}$, while using both $\gamma_3/\gamma_1 =0,0.1$ to model unwarped and warped ABC-MLG, respectively, and also vary the chemical potential through $\mu$ \cite{kaladzhyanSurfaceStatesQuasiparticle2021}. }

{As we are interested in the ABC-MLG surface in the limit of $N\gg1$ layers, we introduce a virtual wall of infinitely large (non-magnetic) impurities $\hat{V}_{\text{imp}}(\boldsymbol{r})=V\delta(z)\hat{1}$, separating the bulk system Eq.~\eqref{eq:Hbulk} into two semi-infinite pieces along the $z$-direction, as illustrated schematically in Fig.~\ref{fig:sm-1}. 
Two surfaces, $S$ and $S'$, are then produced, one for each semi-infinite system, but we will henceforth only consider surface $S$.
Following the $T$-matrix formalism, we can compute a {\it surface} Green's function $\hat{G}(\tilde{\boldsymbol{k}}_{1},\tilde{\boldsymbol{k}}_{2},i\omega_{n})$ from the bulk Green's function $\hat{G}_0(\tilde{\boldsymbol{k}},i\omega_{n})$ obtained from $\mathcal{H}(\tilde{\bs{k}})$, using \cite{Pinon20,kaladzhyanSurfaceStatesQuasiparticle2021} 
$    \hat{G}(\tilde{\boldsymbol{k}}_{1},\tilde{\boldsymbol{k}}_{2},i\omega_{n})  =\hat{G}_{0}(\tilde{\boldsymbol{k}}_{1},i\omega_{n})\delta_{\tilde{\boldsymbol{k}}_{1},\tilde{\boldsymbol{k}}_{2}} +\hat{G}_{0}(\tilde{\boldsymbol{k}}_{1},i\omega_{n})
\hat{T}(\tilde{\boldsymbol{k}}_{1},\tilde{\boldsymbol{k}}_{2},i\omega_{n})
\hat{G}_{0}(\tilde{\boldsymbol{k}}_{2},i\omega_{n})$
with $\hat{T}=V\delta_{k_{1x},k_{2x}}\delta_{k_{1y},k_{2y}}[\hat{1}-V\sum_{k_{z}} \hat{G}_{0}(k_{1x},k_{1y},k_{z},i\omega_{n})]^{-1}$ and $\omega_{n}$ the Matsubara frequency.
Finally, we extract the surface Green's function for the S surface adjacent to the impurity plane, $\hat{\mathcal{G}}(\boldsymbol{k},i\omega_{n})$, by partial Fourier transform of $\hat{G}(\tilde{\boldsymbol{k}}_{1},\tilde{\boldsymbol{k}}_{2},i\omega_{n})$ in the $z$-direction, using henceforth the in-plane notation $\boldsymbol{k}=(k_x,k_y)$. We here compute the Matsubara summations using the effective Ozaki's summation \cite{ozakiContinuedFractionRepresentation2007b}.
We also perform the analytic continuation $i\omega \to \omega +i \eta$ ($\eta=10^{-4}$) for the bulk Green's function. 
This results in an effective surface Green's function of a semi-infinite ABC-MLG stack with a total of six sublattice (carbon) sites in the surface single unit cell, i.e.~the surface unit cell is three layers deep, giving us also the depth profile, for the three atomic layers closest to the surface.}

{
Finally, since two surfaces are left by the impurity wall, but we only access one, our states must be replicated for the other surface. Equivalently, for any finite system, the slab will also have two surfaces.
However, as shown in Fig.~\ref{fig:sm-1}, one needs to exchange the results as $B_3\rightarrow A'_{1}$ and $A_3\rightarrow B'_{1}$ to get the correct results. In other words, the top surface $A$ ($B$) gets mapped to the bottom surface  $B$ ($A$) with the exact opposite magnetic moment.  Such magnetization pattern across the two surface has been coined a layer antiferromagnetic (LAF) state \cite{Macdonald2011,hanCorrelatedInsulatorChern2024}. This refers to the fact that the net magnetic moment of the top-most surface, comprising of $B_3,~A_3$ atoms in Fig.~\ref{fig:sm-1}, is exactly opposite to that of the bottom-most surface, comprising of the $B'_1,~A'_1$ atoms in. As a consequence, the bottom-most and top-most layers are antiferromagnetically correlated.} 

{
Another way to understand this mapping and also why the LAF state appear very naturally in ABC-MLG, beyond the somewhat technical impurity wall construction above, is to consider a thick ABC-MLG slab with a top and bottom surface. On each surface only one sublattice is participating in the flat band, and, by symmetry, the sublattice changes between the two surfaces (this is directly visible from analyzing a finite slab consist of unit cells of Eq.~\eqref{eq:Hbulk}).  From our, and also earlier results, including first-principles results, we also know that the sublattice carrying the flat band is also the one carrying the majority of the magnetization. 
Thus, if, say the $A$ sublattice makes up the flat band on the top surface and thus hosts the majority of the magnetization on the top surface, say spin-up polarization, then the $B$ sublattice hosts the majority of the magnetization on the bottom surface.
However, no real system is infinitely thick, so there will always be some cross-interactions between the two surfaces. In order to avoid a domain wall in the middle of the slab, which always costs energy, the magnetization needs to have the same sign on each sublattice throughout the whole slab.  Thus, the $A$ sublattice will always have spin-up polarization, while the $B$ sublattice will always have spin-down polarization. Hence the magnetization on the bottom surface is both opposite that of the top surface and sits on the opposite sublattice.}

\begin{figure}[ht]
\begin{center}
\includegraphics[width=0.5\linewidth]{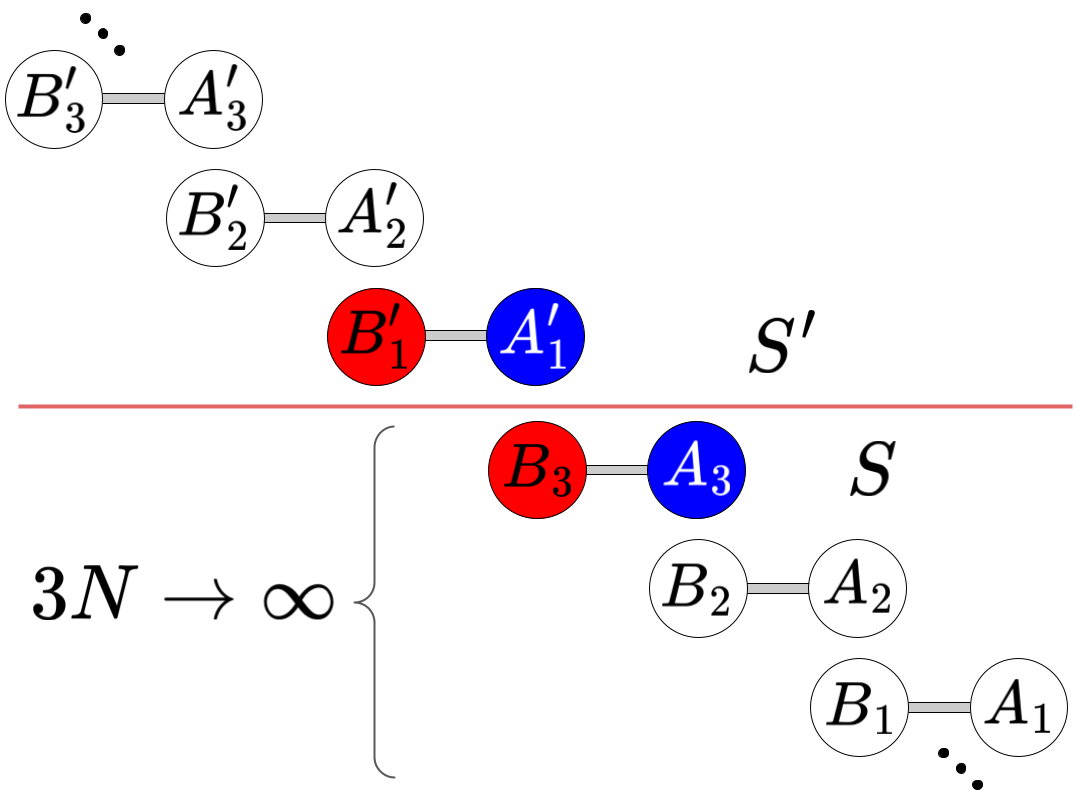}
\end{center}
\caption{{Illustration of the impurity wall method used for ABC-MLG \cite{kaladzhyanSurfaceStatesQuasiparticle2021}. Different color represent opposite spin magnetization direction. Two surfaces, $S$ and $S'$, are left behind, one for each semi-infinite system, although we only access $S$. Any magnetic states we find, therefore, have to be replicated for the other surface, by exchanging the results as $B_3\rightarrow A'_{1}$ and $A_3\rightarrow B'_{1}$, while keeping the magnetization direction constant on each sublattice since they ultimately belong to the same system. These are thus always layer antiferromagnetic (LAF) states. }}
\label{fig:sm-1}
\end{figure}

\section{Static susceptibility profile around the high-symmetry points}
\label{Sec:SI0}

In the main text in Figs.~2 and 3 we investigate the behavior of the static susceptibility $\chi_s'(\bs{q},\omega=0)$ only along a high-symmetry BZ line. 
In this section, we provide additional data on the momentum profile of the static susceptibility motivating this choice for the half-filling case, while also demonstrating that it is not enough at arbitrary doping. 
We do so by examining the microscopic structure around the high-symmetry points ${\bs q}={\bs \Gamma}$ and ${\bs K}$ for the homogeneous spin susceptibility and the most dominating density-density element $B_3 B_3$ in $\hat{\chi}_0'(\bs{q},\omega=0)$ at half-filling in Fig.~\ref{SM_chi_map} and away from half-filling in Fig.~\ref{SM_bare_chi_map}, respectively.

At half-filling we find that the static spin susceptibility $\chi'_s(\bs{q},\omega=0)$ diverges over a ring-like region around the ${\bs q}={\bs \Gamma}$ and ${\bs K}$, see Fig.~\ref{SM_chi_map} upper (lower) panel for ${\bs q}={\bs \Gamma}$ (${\bs q}={\bs K}$). 
While the rings have a hexagonal (trigonal) distortion, they still respect the $C_6$ ($C_3$) rotational symmetry of the lattice, and the divergences are spread uniformly along the ring.
Upon increasing $U$, the ring expands but maintains the hexagonal (trigonal) symmetry. 
Overall, this analysis suggests the divergent nesting vectors move away from the commensurate nesting vectors ${\bs \Gamma}$ and ${\bs K}$ as only obtained for $U=0$.  Still, by keeping the hexagonal (trigonal) lattice symmetry with a uniform divergence along the perimeter, all scattering vectors with divergent susceptibility can be captured by only examining the high-symmetry line $\Gamma\to M \to K\to \Gamma$. 
This validates the choice of ${\bs k}$-values (along the $x$-axis) in Figs.~2 and 3.

Next, we add a finite doping which causes the Fermi level to move away from the flat band region. 
We examine the static bare susceptibility $\chi'_0(\bs{q},\omega=0)$ as a function of finite doping $\mu$ in Fig.~\ref{SM_bare_chi_map}.
At finite doping we find a star-like divergence instead of a ring divergence around ${\bs q}={\bs \Gamma}$ point, see upper panel of Fig.~\ref{SM_bare_chi_map}, while it continues to form a triangular shape around  ${\bs q}={\bs K}$, see lower panel of Fig.~\ref{SM_bare_chi_map}. Notably, the degree of divergence is not uniformly distributed over the star-like or triangle-like regions, rather there exist certain points yielding the strongest divergence. As a result, the equidistant nature of the divergent scattering momentum with respect to the high-symmetry points is no longer present. This means that the high-symmetry line through the BZ might not contain the strongest divergences.  As a result, to capture the $U_c$ for the doped case, we need to go beyond the high-symmetry line such that we do not miss the correct order. We do this in Fig.~4 by extracting $U_c^*$ over the whole disc centered around ${\bs \Gamma}$ and ${\bs K}$. We also note that the regions with star-like and triangle-like divergences expand with increasing doping, resulting in an outward shift in incommensurate nesting vectors with respect to the high-symmetry points, as also illustrated in the inset in Fig.~4. 

As a naming convention, we refer to the incommensurate nesting vectors lying on the high-symmetry line as ${\bs K}'$ and ${\bs \Gamma}'$ in the case of half-filling. On the other hand, we adopt the ${\bs K}''$ and ${\bs \Gamma}''$ notation for the incommensurate nesting vectors in the case of finite doping that can lie both away from as well as on the high-symmetry line.


\begin{figure}[ht]
\begin{center}
\includegraphics[width=1\linewidth]{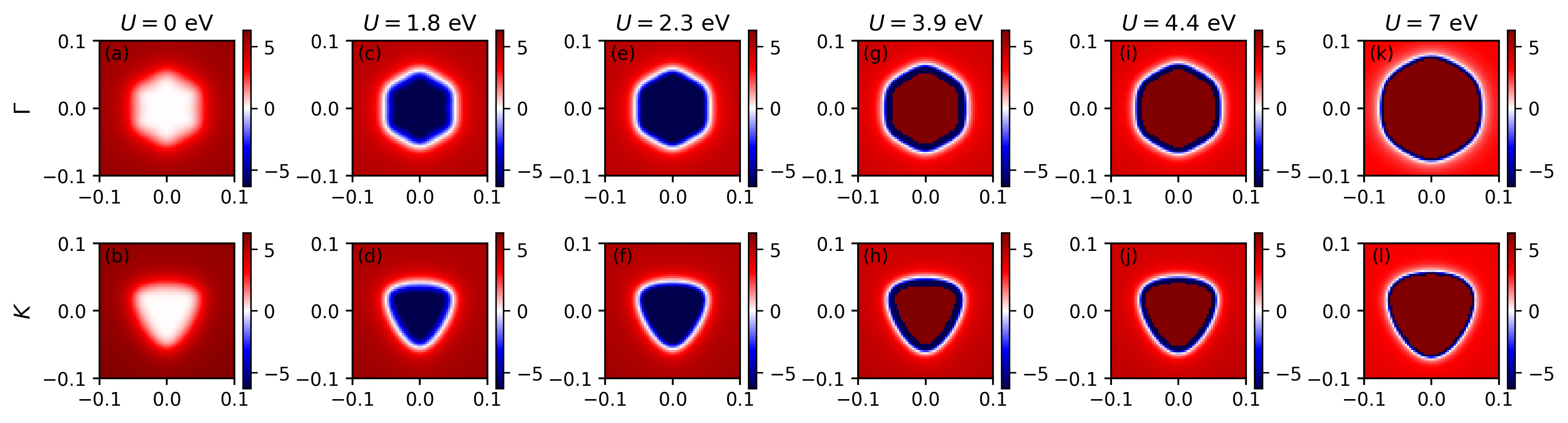}
\end{center}
\caption{Inverse of the homogeneous static spin susceptibility $(\chi'_s(\bs{q},\omega=0))^{-1}$ around $\bs{\Gamma}$ (first row) and $\bs{K}$ (second row) high-symmetry points at half-filling, $\mu=0$. Blue color denotes negative values, red denotes positive values, and white denotes divergent susceptibility values. The $U=0$ values cut off colors.}
\label{SM_chi_map}
\end{figure}


\begin{figure}[ht]
\begin{center}
\includegraphics[width=1\linewidth]{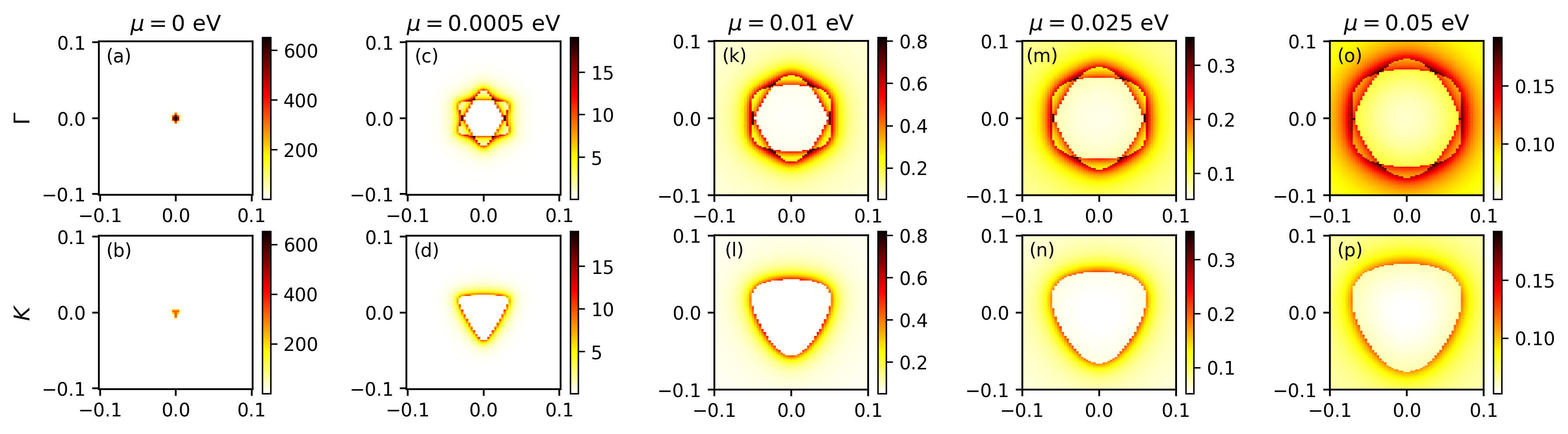}
\end{center}
\caption{The real part of  the static spin susceptibility $\chi_0(\boldsymbol{q},\omega=0)$ around $\bs{\Gamma}$ (first row) and $\bs{K}$ (second row) high-symmetry points at fillings $\mu=0,0.0005,0.01,0.025$, and $0.05$~eV.  }
\label{SM_bare_chi_map}
\end{figure}

\clearpage

\section{Dynamic spin susceptibility analysis for zero and small $U$ at half-filling}
\label{Sec:SI1b}

In the main text we discuss the role of flat bands in forming magnetic order in the non-interacting system in the section ``\textit{Flat band and nesting effects}". Here we provide additional data supporting those results. In particular, we present the full frequency behavior of the real part of the spin susceptibility $\chi^{\prime}_{s}(\bs{q},\omega)$ (dashed lines) and the imaginary part of the spin susceptibility $\chi^{\prime \prime}_{s}(\bs{q},\omega)$ (solid lines)  for zero and small values of $U$ at the commensurate nesting vectors ${\bs q}={\bs \Gamma}$ (black), ${\bs K}$ (red) in Fig.~\ref{SM_fig3}.

\begin{figure}[ht]
\begin{center}
\includegraphics[width=0.98\linewidth]{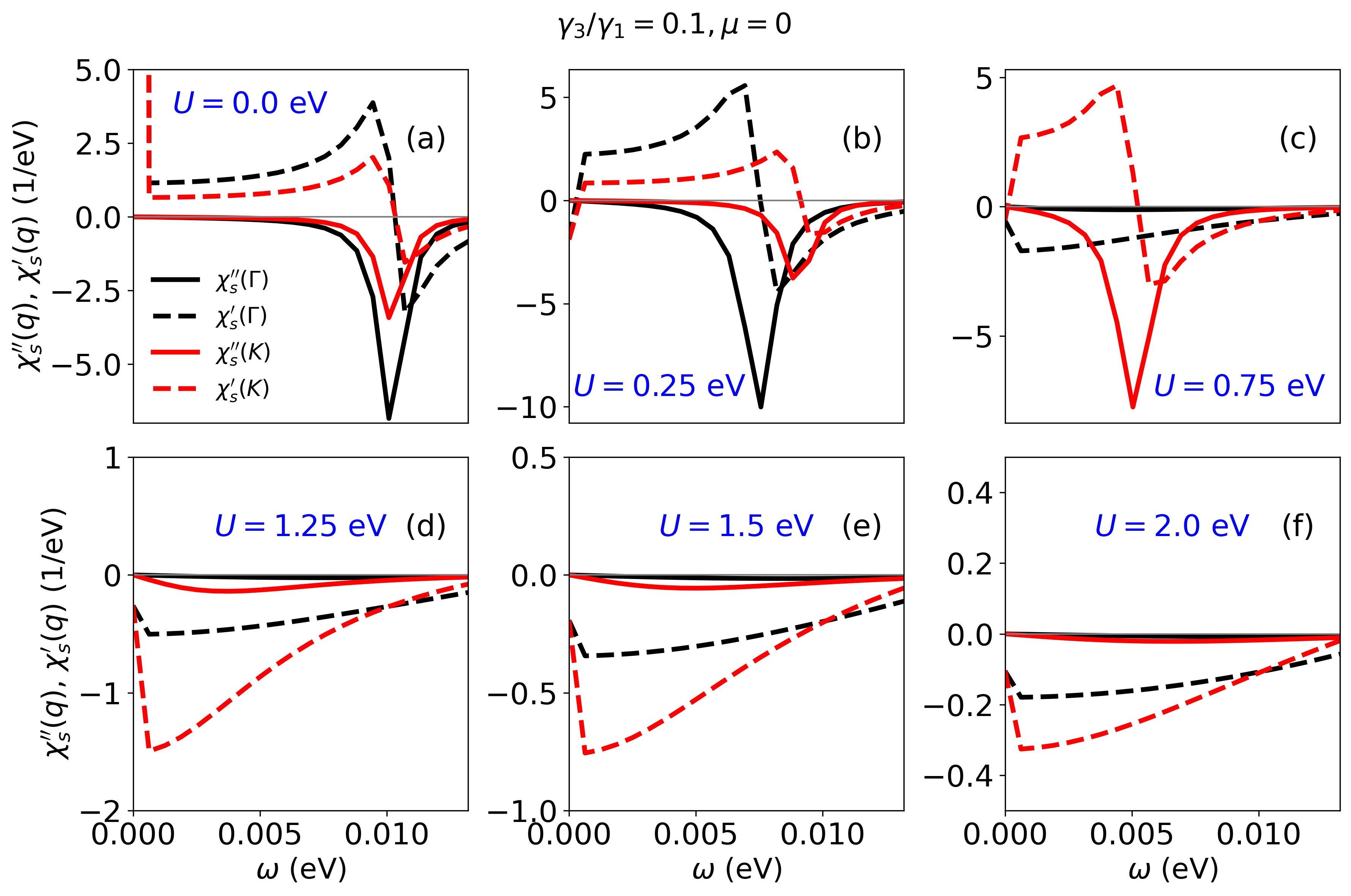}
\end{center}
\caption{Evolution of the dynamic spin susceptibility $\chi_{s}(\bs{q},\omega)$ for $\bs{q}=\bs{K}$ (red) and $\bs{\Gamma}$ (black) as a function of frequency $\omega$ for $U=0,0.25,0.75,1.25,1.5$, and $2.0$~eV (a-f), with real part $\chi^{\prime}_{s}(\bs{q},\omega)$  (dashed) and imaginary part $\chi^{\prime \prime}_{s}(\bs{q},\omega)$ (solid). Here $\gamma_3/\gamma_1=0.1$ and $\mu=0$~eV, while $\bs{K}\approx(2.09,1.21)$ and $\bs{\Gamma}=(0,0)$.
}
\label{SM_fig3}
\end{figure}

Before proceeding, we note that the susceptibility accounts for the probability of the scattering taking place with a transfer of momentum ${\bs q}$ and energy $\omega$. In general, the real part of the spin susceptibility $\chi_s^\prime$ estimates how much the system favors a particular magnetic order possibility, while the imaginary part $\chi_s^{\prime \prime}$ encodes the loss in the system. We note that the gap along the frequency axis does not necessarily correspond to a spectral gap, rather it represents the energy barrier for opposite spin correlations.  Therefore, the frequency where the susceptibility peaks can be considered to mark a spin gap associated with the magnetic order \cite{bruusManyBodyQuantumTheory2004}.  

Starting from $U=0$ in Fig.~\ref{SM_fig3}(a), we find a (negative) resonance peak at finite $\omega$ in the imaginary part of spin susceptibility for both ${\bs q}={\bs K}$ and ${\bs \Gamma}$.
Exactly at the same values of $\omega$, the real parts exhibit a discontinuous jump. A closer inspection suggests that the real part changes its sign at a zero crossing where the imaginary part shows the resonance peak. 
This marks the existence of an ordered state with a finite spin gap for both intra- and inter-valley scattering already at $U=0$. 
We also note that $\chi^{\prime}({\bs q},\omega=0)$ changes its sign from positive to negative as soon as an infinitesimal $U$ is considered.  
These features continue to exist for finite but small $U$, except that the resonance peaks move steadily towards zero frequency when $U$ increases, see Fig.~\ref{SM_fig3}(b). 
Once $U$ crosses $U^{-}_{\rm c} \approx 0.5$~eV, the resonance peak as well as discontinuous jump for ${\bs q}={\bs \Gamma}$ are not visible anymore and the ordering at ${\bs \Gamma}$ is thus lost, see Fig.~\ref{SM_fig3}(c). 
The same occurs for inter-valley scattering at $U>U^{+}_{\rm c} \approx 1.0$~eV, see Fig.~\ref{SM_fig3}(d). 
Thus, beyond this $U_c^+$ ABC-MLG becomes magnetically disordered as far as the ordering vectors ${\bs \Gamma},{\bs K}$ are concerned, see Fig.~\ref{SM_fig3}(e,f). 
We refer to the order at zero and small $U$ as the sFM orders in the main text, where Fig.~2 summarizes the main features in the susceptibility. In Sec.~\ref{Sec:SI4}, we discuss the real-space pattern of the spin moments on the surface of the ABC-MLG. 

Two additional points are worth commenting on. 
First, the critical values $U<U^{\pm}_{\rm c}$ mark the termination of an ordered state. 
This is not determined by the Stoner criterion, which only marks the onset of ordering. 
Nonetheless, the clear vanishing of prominent features in the spin susceptibility, including the vanishing of a clear spin gap for $U>U^{\pm}_{\rm c}$ makes it clear that the preceding order existing at lower $U$ values cannot exist anymore. 
Second, $\chi^{\prime \prime}_{s}(\bs{q}=\bs{\Gamma, K},\omega=0)$ display a negatively valued peak. 
This is the opposite sign compared to the features marking the transitions at $U_c^{\Gamma',K'}$ as shown below in Sec.~\ref{Sec:SI1} and Fig.~\ref{SM_fig1}.  These different signs are required for the spin susceptibility to be continuous across the full $U$ range.
Their existence also marks that the FiM orders occurring at $U_c^{\Gamma',K'}$ must have had a predecessor ordering at lower $U$ values.

\clearpage


\section{Dynamic spin susceptibility analysis for finite $U$ at half-filling}
\label{Sec:SI1}
In the main text, we show the existence of FiM order for realistic strengths of the Hubbard repulsion $U$ in the section ``\textit{FiM orders with interactions}" staying at half-filling. In this section of the SM, we provide additional data for the dynamic spin susceptibility $\chi_{s}(\bs{q},\omega)$ with varying finite $U$. In particular, the analysis is useful to further understand the magnetic phase transitions at $U^{K'}_{\rm c} \approx 2.30$eV and $U^{\Gamma'}_{\rm c} \approx 4.40$eV, established in Figs.~3 of the main text. 
We thus repeat Fig.~\ref{SM_fig3} but now focus on larger values of $U$ and use the incommensurate nesting vectors ${\bs q}={\bs \Gamma}', {\bs K}'$. 
The results are presented in Fig.~\ref{SM_fig1}.

\begin{figure}[h!bt]
\begin{center}
\includegraphics[width=0.98\linewidth]{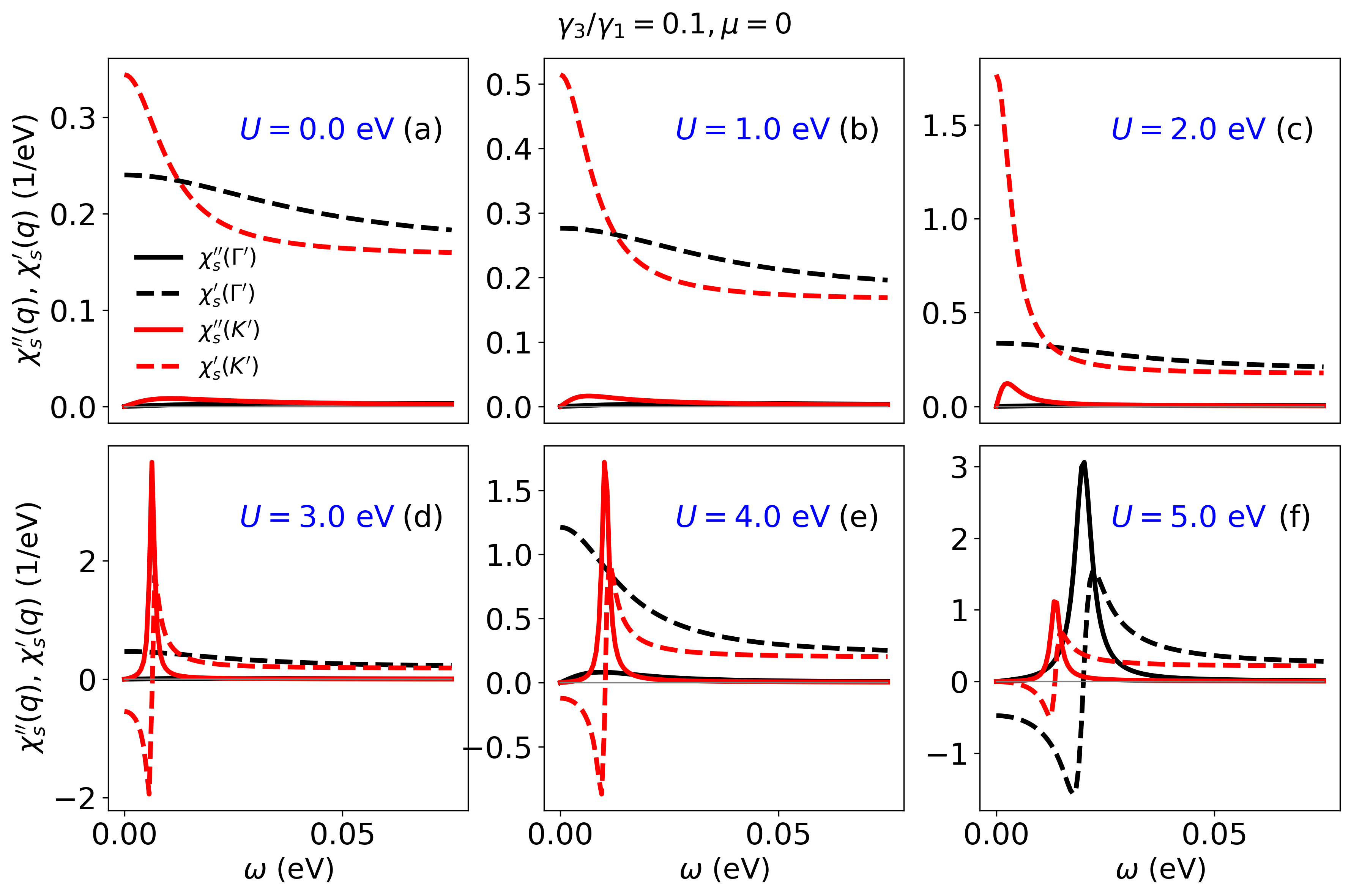}
\end{center}
\caption{Evolution of the dynamic spin susceptibility  $\chi_{s}(\bs{q},\omega)$ for $\bs{q}=\bs{K}'$ (red) and $\bs{\Gamma}'$ (black) as a function of frequency $\omega$ for $U=0,1.0,2.0,3.0,4.0$, and $5.0$~eV (a-f), with real part $\chi^{\prime}_{s}(\bs{q},\omega)$  (dashed) and imaginary part $\chi^{\prime \prime}_{s}(\bs{q},\omega)$ (solid). Here $\gamma_3/\gamma_1=0.1$ and $\mu=0$~eV, while $\bs{\Gamma}'\approx(0.07,0)$ and $\bs{K}'\approx(2.09,1.15)$.
For reference, $\bs{K}\approx(2.09,1.21)$ and $\bs{\Gamma}=(0,0)$.}
\label{SM_fig1}
\end{figure}

For comparison, we start with $U=0$ in Fig.~\ref{SM_fig1}(a) where we find $\chi^{\prime}_{s}(\bs{K}',\omega)$ and $\chi^{\prime}_{s}(\bs{\Gamma}',\omega)$ both decaying with $\omega$ from their $\omega=0$ values. 
On the other hand,  both $\chi^{\prime \prime}_{s}(\bs{K}',\omega)$ and $\chi^{\prime \prime}_{s}(\bs{\Gamma}',\omega)$ start from zero at $\omega=0$ but then acquire a finite but small value very close to $\omega=0$,  before they again decay to zero with $\omega$. 
We hence have no order at ${\bs \Gamma'},{\bs K'}$ at $U=0$, but the order only exists at the commensurate ${\bs \Gamma},{\bs K}$ as seen in Fig.~\ref{SM_fig3}.
With increasing $U$, see Fig.~\ref{SM_fig1}(b,c), $\chi^{\prime}_{s}(\bs{K}',\omega)$ develops more and more of a peak at $\omega = 0$ while $\chi^{\prime \prime}_{s}(\bs{K}',\omega)$ simultaneously develops a peak at small and decreasing $\omega$. 
Such a peak is referred to as a resonance phenomenon since the measure of such dynamic correlation changes with frequency. 
At the same time, no noticeable changes occur for $\chi_s(\bs{\Gamma}',\omega)$ for these $U$ values.  

As soon as $U$ hits $U^{K'}_{\rm c}$, $\chi^{\prime \prime}_{s}(\bs{K}',\omega)$ forms a resonance peak at $\omega=0$, which for $U>U^{K'}_{\rm c}$ is shifted to a finite value of $\omega$, see Fig.~\ref{SM_fig1}(d). 
At the same time, $\chi^{\prime}_{s}(\bs{K}',\omega)$ changes its character from a single positive peak to a sign-changing divergence, or at least a discontinuous jump crossing the zero-line, appearing at $\omega =0$ at the transition and then moving to finite $\omega$ with increasing $U$. The peaks/divergences in $\chi^{\prime}_{s}(\bs{K}',\omega)$ and $\chi^{\prime \prime}_{s}(\bs{K}',\omega)$ are clearly appearing at the same $\omega$ (real part crosses zero when the imaginary part peaks) at and beyond $U^{K'}_{\rm c}$. This exact matching of the real and imaginary spin susceptibility clearly marks the onset of magnetic ordering with ordering vector $\bs{K}'$ at $U^{K'}_{\rm c}$, consistent with many earlier analyses of ordering within RPA calculations in other materials \cite{Christensen16,Fong2000,Gao2010}. 
We refer to this order in the main text as the $\bs{K}'$-FiM state.
The same evolution is also observed in Figs.~\ref{SM_fig3}(a,b,c) for $U<U^{\pm}_{\rm c}$ in the sFM orders.

Moving on to larger $U$ values we find that the peak and discontinuous jumps in $\chi_{s}(\bs{K}',\omega)$ slowly decrease, see Fig.~\ref{SM_fig1}(e). 
Instead, we find that the same features as discussed above appear for an ordered magnetic state at $\bs{\Gamma}'$, setting in at $U^{\Gamma'}_{\rm c}$, see Fig. \ref{SM_fig1}(f).
As a consequence, for $U>U^{\Gamma'}_{\rm c}$, ${\bs \Gamma}'$-FiM order is formed. 
Due to the simultaneous suppression of the features in $\chi_{s}(\bs{K}',\omega)$, we conclude that the ${\bs \Gamma}'$-FiM may dominate at these larger $U$ values. 
This result supports the susceptibility results of Fig.~3 in the main text.
In Sec.~\ref{Sec:SI4}, we discuss the accompanied real-space pattern of the spin moments on the surface of the ABC-MLG. 

\clearpage


\section{Dynamic spin susceptibility analysis with $U$ away from half-filling}
\label{Sec:SI2}


In the main text, we analyze the case of finite doping away from half-filling in the section ``{\it Finite doping}". 
In this section of the SM, we provide additional data for the dynamic spin susceptibility. 
In Fig.~\ref{SM_fig22} we redo the analysis of Figs.~\ref{SM_fig3} and \ref{SM_fig1} for the representative doping level  $\mu=0.023$~eV.
This analysis is useful to understand the underlying data of Fig.~4 of the main text where we plot extracted $U_c^*$ as a function of $\mu$. 
Note that we here are not restricted to a high-symmetry line in BZ when extracting the nesting vectors, but we identify the ${\bs q}$-values with divergent spin susceptibilities in a whole disc around ${\bs q}={\bs K},{\bs \Gamma}$. 
For notational consistency, we adopt ${\bs q}={\bs \Gamma}''$, ${\bs K}''$ representation away from the half-filled cases as already mentioned in Sec. \ref{Sec:SI0}.

\begin{figure}[ht]
\begin{center}
\includegraphics[width=0.98\linewidth]{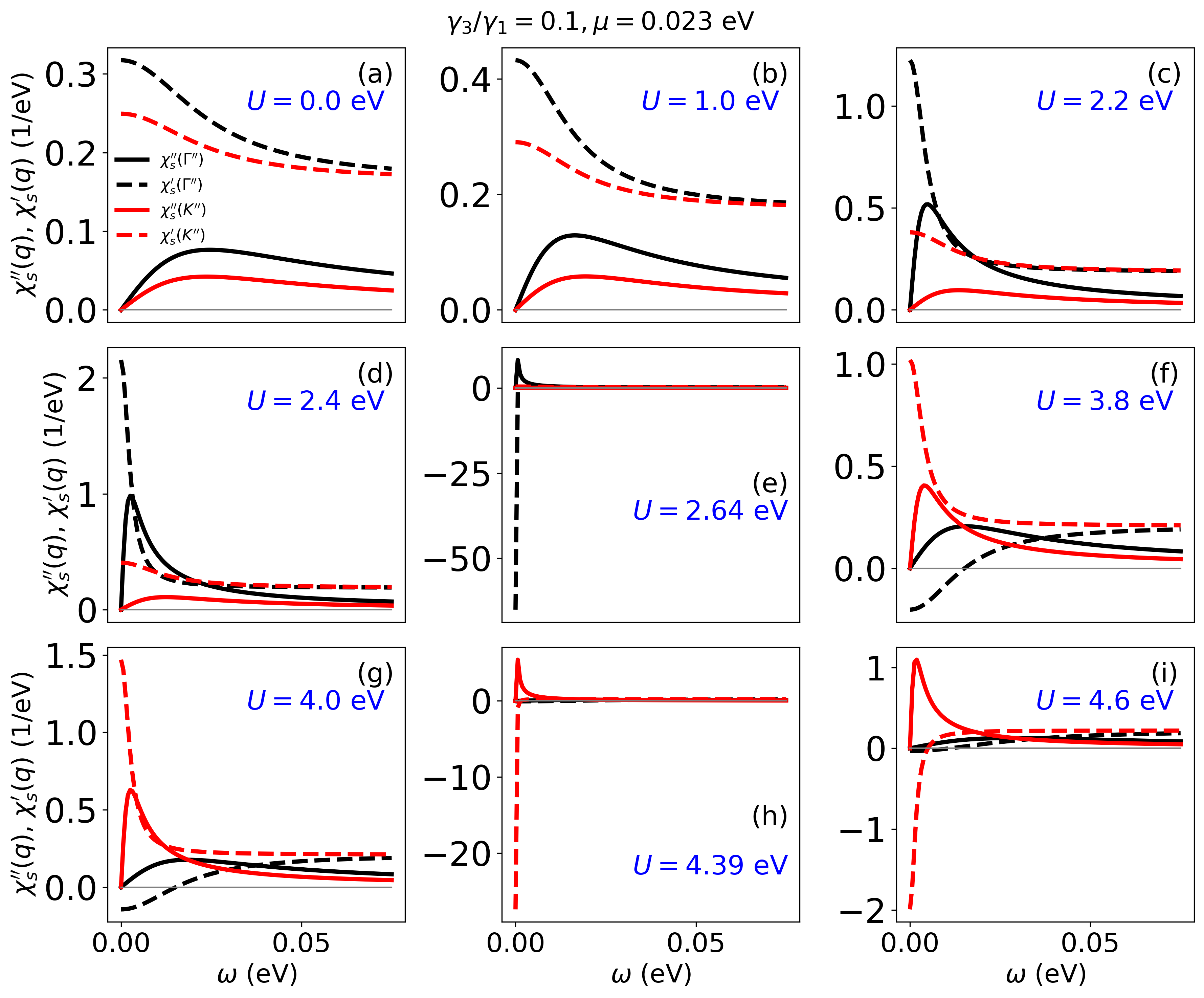}
\end{center}
\caption{
Evolution of the dynamic spin susceptibility  $\chi_{s}(\bs{q},\omega)$ for $\bs{q}=\bs{K}''$ (red) and $\bs{\Gamma}''$ (black) as a function of frequency $\omega$ for $U=0.0, 1.0, 2.2, 2.4, 2.64, 3.8, 4.0, 4.39$, and $4.6$~eV (a-i), with real part $\chi^{\prime}_{s}(\bs{q},\omega)$  (dashed) and imaginary part $\chi^{\prime \prime}_{s}(\bs{q},\omega)$ (solid). Here $\gamma_3/\gamma_1=0.1$ and $\mu=0.023$~eV, while $\bs{\Gamma}''=(-0.06,0.00)$ and $\bs{K}''=(2.11,1.26)$. These nesting vectors are chosen from Fig.~4 for the chemical potential $\mu=0.023$~eV. 
For reference, $\bs{K}\approx(2.09,1.21)$ and $\bs{\Gamma}=(0,0)$.
}
\label{SM_fig22}
\end{figure}

At low interaction strengths $U$, we find qualitatively similar behavior compared to Fig.~\ref{SM_fig1} for the real and imaginary parts of the dynamical spin susceptibility, see Figs.~\ref{SM_fig22}(a,b,c,d).
A positive peak starts forming for the real part close to $\omega =0$, while the imaginary part tends towards a resonance peak once $U$ increases. This trend is somewhat more visible for ${\bs q}={\bs \Gamma}''$ case as compared to ${\bs q}={\bs K}''$.  
We then find that $\chi^{\prime}_{s}({\bs \Gamma}'',\omega=0)$ changes its sign from positive to negative, passing through a negative divergence at $U^*_c=U^{\Gamma''}_{\rm c}=2.64$~eV, see Fig.~\ref{SM_fig22}(e). This marks a transition to the ordered state referred to as ${\bs \Gamma''}$-FiM in the main text.
However, we do not find a clear resonance peak in $\chi^{\prime \prime}_{s}({\bs \Gamma}'',\omega)$ or clear discontinuous jumps in $\chi^{\prime}_{s}(\bs{\Gamma}'',\omega)$ beyond the transition point, as we did in the half-filled case. The peak in $\chi^{\prime \prime}_{s}({\bs \Gamma}'',\omega)$ exists but is not very pronounced, while $\chi^{\prime}_{s}({\bs \Gamma}'',\omega)$ rather smoothly changes its sign from positive to negative values at a finite $\omega$ for $U>U^{\Gamma''}_{\rm c}$, see Figs.~\ref{SM_fig22}(e,f,g). Still, the real part crosses zero at the same finite $\omega$ where the imaginary part exhibits a peak. 
This is similar to the half-filled case, where the zeros of the real part coincide with the resonance peak of the imaginary part. However, altogether this marks notable differences in the spin susceptibility for the ${\bf \Gamma''}$-FiM order at finite doping compared to the ${\bf \Gamma'}$-FiM order at half-filling, in addition to the different nesting vectors.
Similar features are noticed at ${\bs  q}=\bs{K}''$, where $\chi^{\prime}_{s}(\bs{K}'',\omega=0)$ exhibits a negative divergence at $U^*_c=U^{K''}_{\rm c}=4.39$~eV, see Fig.~\ref{SM_fig22}(h), signaling the onsite of the ${\bf K''}$-FiM order. 
For $U>U^{K''}_{\rm c}$ the real part again crosses zero at a finite $\omega$ exactly where the imaginary part exhibits a faint peak, see Fig.~\ref{SM_fig22}(i).

The notably different behavior of the dynamical spin susceptibility beyond the ordering transitions  $U^*_c=U^{\Gamma'',K''}$ at finite doping compared to the half-filing case beyond the $U_c^{\Gamma',K'}$ can be explained by damping, which is substantial in the former case. 
To be precise, the behavior of the dynamic spin susceptibility is connected to the underlying spin-spin relaxation time of the ordered phase. 
The flattened (sharp) nature of the imaginary part and the smooth (discontinuous) zero-energy crossing of the real part in the ordered states is due to a smaller (larger) value of the spin-spin relaxation time for finite (zero) doping \cite{barraElectronSpinResonance2005,yalcinFerromagneticResonance2013}.
The analysis presented in Figs.~\ref{SM_fig1} and \ref{SM_fig22} infers that the FiM order at finite doping has a much shorter spin-spin relaxation time. 
We attribute this behavior to Landau damping associated with the metallic surface state at finite doping, which enhances the spin-spin relaxation rate broadening the magnetic resonance peak as also observed in other RPA studies \cite{maierWeakcouplingTheoryNeutron2023}.
Notably, the co-existing Dirac bulk spectrum, also present at half-filling, does not generate any substantial damping.

Finally, we discuss the spin gap for the ${\bf \Gamma''},{\bf K''}$-FiM orders.
In Figs.~\ref{SM_fig_omega_doping} and \ref{SM_fig_omega_doping_K}, we plot the imaginary part of the dynamical spin susceptibility $\chi^{\prime\prime}_{s}(\bs{q},\omega)$ as a function of the repulsion  $U$ for the nesting vectors that generate the lowest critical Hubbard parameters $U^*_c$ extracted in Fig.~4, i.e.~${\bs q}={\bs \Gamma}''$ and ${\bs K}''$, respectively. Note that ${\bs \Gamma}'',{\bs K}''$ change as a function of doping, as also indicated in the labels. We also mark the resonance peak in the ordered states with black dots, which denotes the extracted spin gap.
At half-filling, panel (a) in both figures, the spin gaps for the $\bs{\Gamma}$-sFM and $\bs{K}$-sFM orders are visible for zero and small $U$, but they close already at $U \le 1$~eV, as discussed in the main text, marking the end of the sFM orders. 
Note that the array of the black dots is nothing but the bright trail in Figs.~2(b,d). The situation changes dramatically as soon as one turns on a finite $\mu$. At finite doping, panels (b-j), there is no bright trail in the imaginary spin susceptibility at finite $\omega$ and low $U$-values. 
Instead, we see a sharp peak, almost circular, region develops at the $U_c^*$ values indicated with thin grey lines. 
There is thus no spin gap or ordering for $U<U^*_c$. 
For $U>U^*_c$ we find a faint peak, primarily visible through the extracted black dots, emanating from the peak region and increasing in energy for all explored $U$ values.
We observe the same qualitative behavior for all values of finite doping considered and at both ${\bs \Gamma}''$-and  ${\bs K}''$ nesting vectors.
This suggests that once the ${\bs \Gamma}''$-and  ${\bs K}''$-FiM orders are established at finite $U$, they remain with a finite spin gap. 
This is markedly different from the half-filled scenario where flat band-mediated commensurate sFM orders exist at zero $U$ but are terminated at finite $U$. 
As such, the $\bs{\Gamma}'', {\bs K}''$-FiM orders at finite doping are more similar to the incommensurate $\bs{\Gamma}', {\bs K}'$-FiM orders on-setting at finite $U$ and not the $\bs{\Gamma}, {\bs K}$-sFM orders originating at zero $U$, despite the discrepancy in $U$ values between the former pair.

These above results all support
the results of Fig.~4 in the main text. In Sec.~\ref{Sec:SI4}, we discuss the accompanied real-space pattern of the spin moments on the surface of the ABC-MLG.


\begin{figure}[ht]
\begin{center}
\includegraphics[width=1\linewidth]{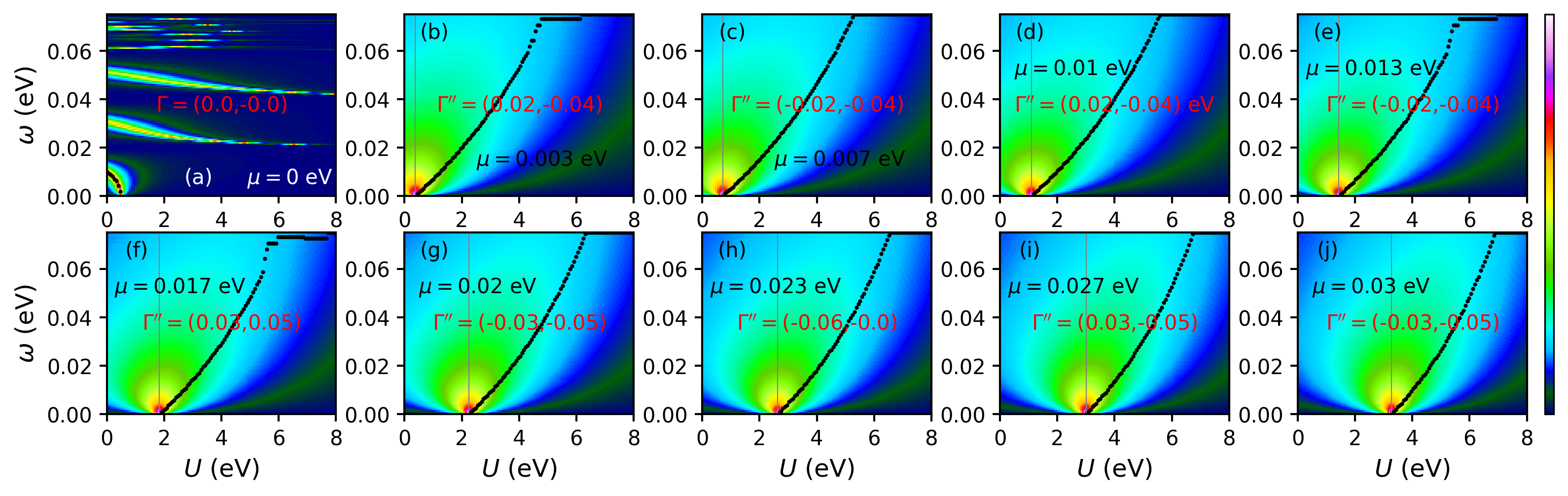}
\end{center}
\caption{Imaginary spin susceptibility $\chi^{\prime \prime}_s({\bs q}, \omega)$ in the $\omega$-$U$ plane for the ordering vector ${\bs q}$ around the ${\bs \Gamma}$-point as a function of increasing doping (a-f). The ordering vector ${\bs \Gamma}''$ varies with doping and is indicated in each panel. 
Vertical thin grey lines denote extracted $U_c^*$ values, also plotted in Fig.~4.
Black dots denote the position of the resonance peak calculated numerically from the data shown in colors. 
}
\label{SM_fig_omega_doping}
\end{figure}


\begin{figure}[ht]
\begin{center}
\includegraphics[width=0.98\linewidth]{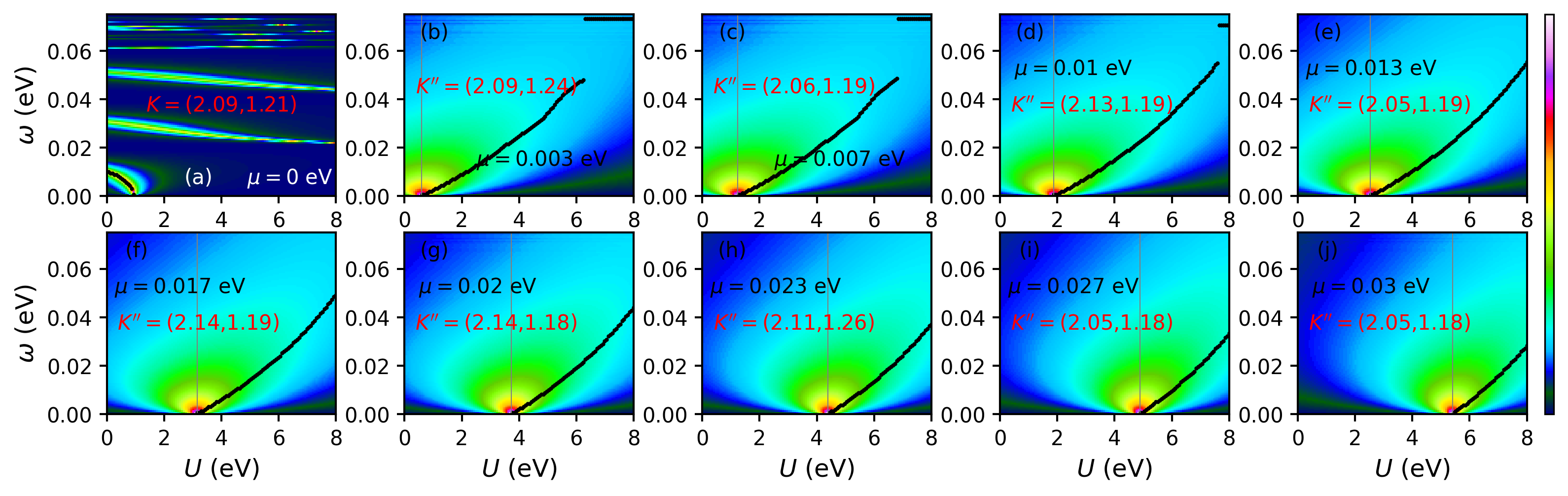}
\end{center}
\caption{Same as Fig.~\ref{SM_fig_omega_doping} but for ordering at ${\bs q}=\bs{K}''$ around the $\bs{K}$-point. 
}
\label{SM_fig_omega_doping_K}
\end{figure}

\clearpage


\section{Magnetic spin textures}
\label{Sec:SI4}

In the main text we discussed and reported data on the magnetic textures, or structure for the sFM orders at half-filling zero and low $U$ in section ``\textit{Flat bands and nesting}" and the FiM states in section ``\textit{FiM order with interactions}" for half-filling and in section ``\textit{Finite doping}" away from half-filling. In this SM section, we provide additional information supporting those results and naming conventions. We can access the magnetic textures by investigating the orbital resolved magnetic moment from the eigenvector, $\bs{\psi}^{\omega=0}_{\bs {q}}$, associated with the static susceptibility matrix $\hat{\chi}_{s}(\bs{q},\omega=0)$ at its divergence, which marks the transition to magnetic ordering \cite{boehnkeSusceptibilitiesMaterialsMultiple2015}.  
We note, however, that these density-density elements in the eigenvector only carry information about the relative distribution of the magnetic moment across the orbitals in a normalized fashion, while the determination of the exact value of the magnetic moment is not straightforward \cite{boehnkeSusceptibilitiesMaterialsMultiple2015,Christensen16,Fong2000,Gao2010}. Below we start by extracting the magnetic texture in the lattice unit cell (here the surface unit cell) and then comment on the real space texture beyond that.

In Fig.~\ref{SM_fig4} we report the distribution of $\bs{\psi}^{\omega=0}_{\bs {q}}$ over the $N_b=6$ orbitals (carbon sites) in the effective surface unit cell we use for our calculations, where $B_3$ and $A_3$ are the atoms of the very top graphene layer. 
As we find that the imaginary parts of these elements are identically zero, we thus left with the real part that we display.
In Fig.~\ref{SM_fig4}(a) we study the sFM states at $U=0$ and half-filling and therefore just use the bare static susceptibility at half-filling. 
From the individual breakdown of the density-density elements in $\psi^{\omega=0,j}_{\bs {q}}$, we find that except the element $B_3 B_3$, all other elements vanish. 
This generates what we refer to as a sublattice ferromagnetic (sFM) order in the very top graphene layer as there exists no opposite spin moment from $A_3 A_3$ element and no moments on any subsurface graphene layers either. 
In Fig.~\ref{SM_fig4}(b) we study the FiM states appearing at finite $U>U^{\Gamma',K'}_{\rm c}$ at half-filling, again plotting the density-density elements in $\psi^{\omega=0,j}_{\bs {q}}$ as a function of the six orbitals in the lattice unit cell.
Here we find that both the $B_3 B_3$  and $A_3 A_3$ elements show finite values, with positive and negative intensities, respectively. 
This marks an antialignment behavior of the moments and due to the strong sublattice polarization, we call this a ferrimagnetic (FiM) order.
We repeat the above analysis for doping away from half-filling using the representative values of $\mu=0.01$~eV and $0.023$~eV in Figs.~\ref{SM_fig4}(c,d), respectively. 
Note that we here must consider different nesting vectors ${\bs q}={\bs \Gamma}''$ and ${\bs K}''$ for the different values of $\mu$. 
Again, the $B_3B_3$ element shows a substantially high value, but also the $A_3A_3$ element is non-zero in both cases and increases with increasing doping. 
We therefore also refer to these as FiM states, although we note that they are structurally relatively similar to the sFM states as well. 
Further note that the orderings in each panel in Fig.~\ref{SM_fig4} host different incommensurability and that the ${\bs K}$-orders also have an extended unit cell.


\begin{figure}[ht]
\begin{center}
\includegraphics[width=0.8\linewidth]{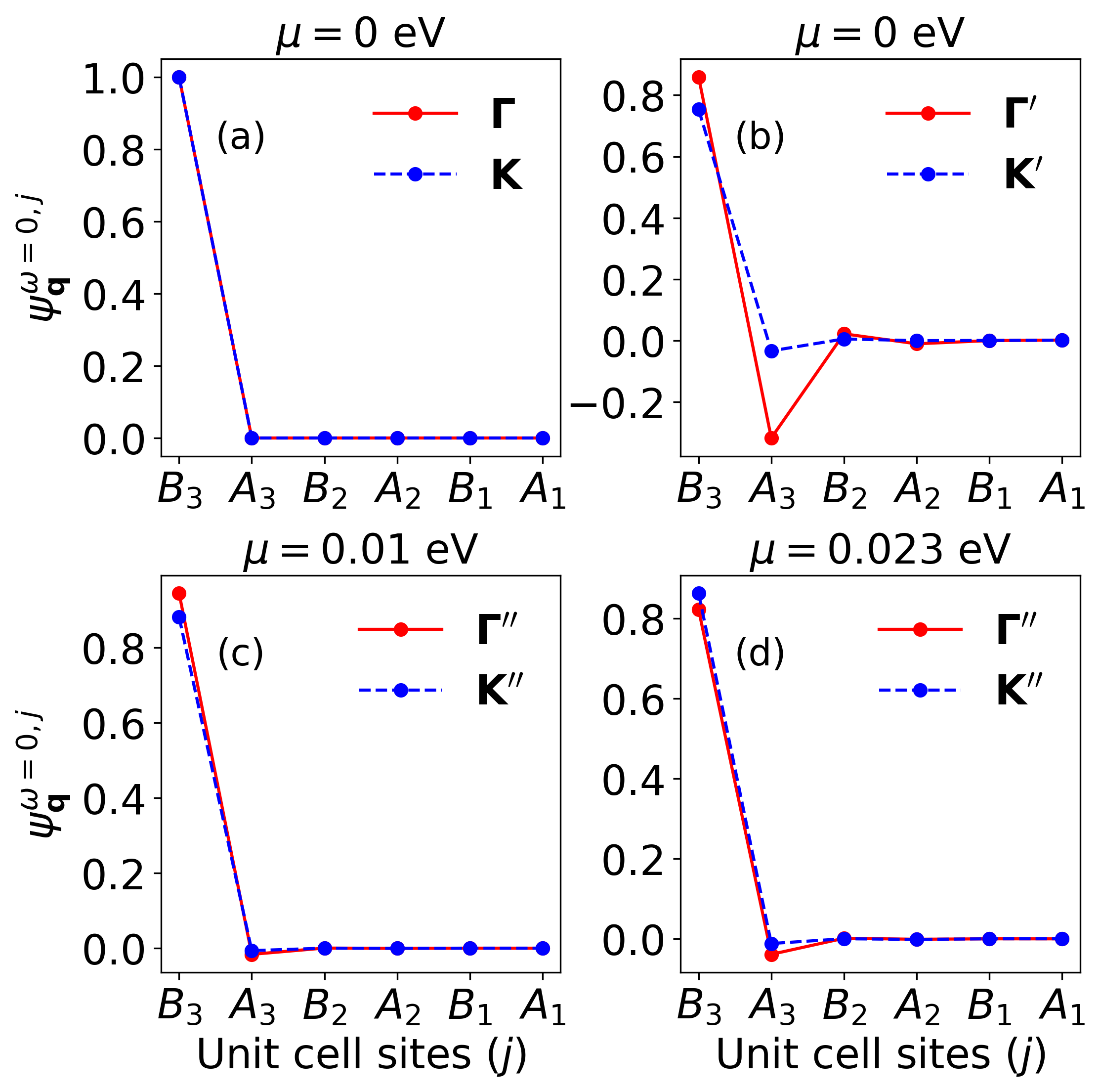}
\end{center}
\caption{Density-density elements in the eigenvector $\psi^{\omega=0,j}_{\bs {q}}$ at each orbital (carbon site) in the ABC-MLG lattice unit cell associated with the divergent static spin susceptibility $\chi_{s}(\bs{q},\omega=0)$ for ${\bs q}={\bs \Gamma}$ (solid red) and ${\bs K}$ (dashed blue) at $U = 0$ and $\mu=0$ (a), ${\bs q}={\bs \Gamma'}$ (solid red) and ${\bs{K}'}$ (dashed blue) at $U = U_c^{\Gamma',K'}-\epsilon$ and $\mu=0$ (b), ${\bs q}={\bs \Gamma''}$ (solid red) and ${\bs{K}''}$ (dashed blue) at $U = U_c^{\Gamma'',K''}-\epsilon$ and $\mu=0.01$~eV (c), and ${\bs q}={\bs \Gamma''}$ (solid red) and ${\bs{K}''}$ (dashed blue) at $U = U_c^{\Gamma'',K''}-\epsilon$ and $\mu=0.023$~eV (d). Here $\epsilon \ll 1$, such that $U$ is chosen right before the critical transition to probe the main fluctuations resulting in the spatial structure of the static order.
The scattering vectors are chosen to match the peaks in Figs.~2(a,c), 3(a,c), and Fig.~4. 
In (a) $\bs{K}\approx(2.09,1.21)$ and $\bs{\Gamma}=(0,0)$, in (b) $\bs{\Gamma}'\approx(0.10,0)$ and $\bs{K}'\approx(2.09,1.13)$, in (c) $\bs{\Gamma}''=(0.02,-0.04)$ and $\bs{K}''=(2.13,1.19)$, and in (d) $\bs{\Gamma}''=(-0.06,0.00)$ and $\bs{K}''=(2.11,1.26)$.}
\label{SM_fig4}
\end{figure}

Finally, we extract the full spatial dependence of the magnetic spin texture. 
For the ${\bs \Gamma}$-centered orders this magnetic texture extracted in Fig.~\ref{SM_fig4} gives the complete texture as that order is just repeated in each unit cell. 
For the ${\bs \Gamma}^{',''}$-orders we need to also consider the incommensurability, which adds real-space modulation to the order. 
Due to the small amount of incommensurability, this modulation will be very long-range. 
In Fig.~2(d) in the main text we plot the spatial magnetic texture of the ${\bs \Gamma}^{'}$-FiM order, ignoring the incommensurability which due to its long-range modulation is not noticeable on the displayed length scale.

For the $\bs{K}$-order, the scattering ${\bs q}$-vector dictates an extended $\sqrt{3} \times \sqrt{3}$ unit cell. 
We can numerically extract the magnetic texture at all sites in the lattice by taking the real part of the Fourier-transformed density-density elements $\psi_{\bs{q}}^{\omega=0,j}$ of the susceptibility eigenvector on the top surface ($j={B_3B_3,A_3A_3}$) of ABC-MLG:
\begin{equation}
    \psi_{\bs{r}}^{\omega=0,j}= \psi^{\omega=0,j}_{\bs{q}}e^{i\bs{q}\cdot({\bs{r}-\bs{r}_j})},
\label{eq:SM_psi}
\end{equation}
where $\bs{r}_j$ is the position of the $j$ sublattice site in the (original) lattice unit cell. 
Applying Eq.~\eqref{eq:SM_psi} we find that all $\bs{q}=\bs{\Gamma}$ orders trivially repeat with periodicity of the lattice, while the $\bs{q}=\bs{K}$ orders result in a $\sqrt{3} \times \sqrt{3}$ extended unit cell. 
{Numerically, we find that the sum of the spin densities in the $\sqrt{3} \times \sqrt{3}$ extended unit cell is  zero, but with time-reversal symmetry broken, this still suggests ferrimagnetic (or an altermagnetic) order, and not antiferromagnetic order.}
The previously shown dynamic susceptibility maps do not present the altermagnet characteristic two-peaked structure due to the chiral magnons \cite{maierWeakcouplingTheoryNeutron2023,smejkalChiralMagnonsAltermagnetic2023}, establishing the $\bs{K}$ order as zero-magnetization ferrimagnetic state.
At realistic $U$ and finite doping, incommensurate nesting breaks the perfect zero net magnetization, transforming these states into finite-magnetization FiM states.
In the inset of Fig.~3(b) in the main text we plot the resulting magnetic texture for the $\bs{K}'$-FiM order. 
Again, we here ignore the slight incommensurability which would generate a long-range additional modulation not visible on the length scales plotted in Fig.~3. 
We here only consider the real part of Eq.~\eqref{eq:SM_psi} to maintain the $\psi_{\bs{r}=0}^{\omega=0,j}=\psi^{\omega=0,j}$ condition, which is required for the lattice unit cell.



\end{onecolumngrid}

\end{document}